\documentclass[aps,prd,twocolumn,sgroupedaddress]{revtex4-2}

\usepackage{amssymb}
\usepackage{amsmath}
\usepackage{graphicx}
\usepackage{enumerate}
\usepackage{mathtools}
\usepackage{color}
\usepackage{bbold}
\usepackage{subfigure}
\usepackage{slashed}
\usepackage[dvipsnames,usenames]{xcolor}
\usepackage{bm}
\usepackage{ulem}

\begin{document}


\title{Collisional instabilities of neutrinos and their interplay with fast flavor conversion in compact objects}


\author{Lucas Johns}
\email[]{ljohns@berkeley.edu, NASA Einstein Fellow}
\affiliation{Departments of Astronomy and Physics, University of California, Berkeley, CA 94720, USA}

\author{Zewei Xiong}
\email[]{z.xiong@gsi.de}
\affiliation{GSI Helmholtzzentrum f\"{u}r Schwerionenforschung, Planckstra{\ss}e 1, 64291 Darmstadt, Germany}

\begin{abstract}
Fast and collisional flavor instabilities possibly occur in the neutrino decoupling regions of core-collapse supernovae and neutron-star mergers. To gain a better understanding of the relevant flavor dynamics, we numerically solve for the collisionally unstable evolution in a homogeneous, anisotropic model. In these calculations collisional instability is precipitated by unequal neutrino and antineutrino scattering rates. We compare the solutions obtained using neutral-current and charged-current interactions. We then study the nonlinear development of fast instabilities subjected to asymmetric scattering rates, finding evidence that the onset of collisional instability is hastened by fast oscillations. We also discuss connections to other recent works on collision-affected fast flavor conversion.
\end{abstract}

\maketitle

\section{Introduction}

A bevy of studies from the past several years \cite{sawyer2005, sawyer2009, sawyer2016, chakraborty2016b, chakraborty2016c, tamborra2017, dasgupta2017, wu2017, wu2017b, dasgupta2018, dasgupta2018b, airen2018, abbar2018, abbar2019, capozzi2019b, yi2019, abbar2019b, azari2019, nagakura2019, chakraborty2020, martin2020, johns2020, johns2020b, bhattacharyya2020, abbar2020, shalgar2020b, capozzi2020, xiong2020, shalgar2020, bhattacharyya2020b, morinaga2020, glas2020, abbar2020b, padilla2020, george2020, johns2021, nagakura2021, nagakura2021b, nagakura2021c, padillagay2022, roggero2022, abbar2022, harada2022, dasgupta2022} support the following conclusions: Neutrino radiation fields are susceptible to fast flavor instabilities, which grow on temporal and spatial scales roughly given by $G_F n_\nu$ with Fermi constant $G_F$ and neutrino number density $n_\nu$; a certain type of crossing between the neutrino and antineutrino angular distributions is a necessary and sufficient condition for fast instabilities occurring; and such crossings probably exist at various locations in core-collapse supernovae and neutron-star mergers, notably including the neutrino decoupling regions in those sites.

Therefore, as neutrinos decouple from the background medium, their flavor states are potentially exposed to fast instabilities and collisions operating simultaneously. Recognition of this possibility has inspired an effort to understand fast flavor conversion (FFC) in the presence of collisions \cite{capozzi2019, shalgar2019, martin2021, sigl2022, shalgar2021, kato2021, sasaki2021, hansen2022, johns2022, shalgar2022, shalgar2022b, kato2022}. These works can be regarded more broadly as part of an enterprise to determine how collisions affect oscillation outcomes---or, depending on the viewpoint, how oscillations affect collisional outcomes---in the early universe and compact objects \cite{stodolsky1987, savage1991, sigl1993, mckellar1994, bell2001, lunardini2001, dolgov2002, abazajian2002, wong2002, mangano2002, mangano2005, strack2005, pastor2009, gava2010, mangano2011, mangano2012, castorina2012, cherry2012, sarikas2012b, vlasenko2014, volpe2015, kartavtsev2015, blaschke2016, grohs2016, grohs2017, barenboim2017, johns2016, johns2018, cirigliano2018, richers2019, cherry2020, froustey2021}.

The generic expectation is that collisions counteract flavor transformation. Conversion occurs through the development of coherence, but collisions typically cause decoherence. Contrary to this expectation, and to the findings of other papers \cite{martin2021, sigl2022}, several publications \cite{shalgar2021, kato2021, sasaki2021, hansen2022} have reported scattering-\textit{enhanced} FFC. Ref.~\cite{johns2022} explained the enhancement effect as being a product of homogeneous modeling. The calculations of Refs.~\cite{shalgar2021, kato2021, sasaki2021, hansen2022} evolved initially anisotropic neutrino flavor distributions. In a homogeneous setting, however, distributions inevitably isotropize due to scattering. The calculations consequently exhibit transient evolution as the distributions come to be isotropic.

Pursuing a separate line of thinking, Ref.~\cite{johns2021b} proposed a distinct mechanism---collisional instability---by which collisions might in fact enhance flavor conversion. The study also explained why collisional instability is not in conflict with the idea that collisions cause decoherence. The key is that neutrinos and antineutrinos are decohered differently, and that the differential effect feeds into the nonlinear part of the Hamiltonian. This understanding fits a pattern linking instability types to particular asymmetries between neutrinos and antineutrinos:
\begin{gather}
\notag \\
\textbf{Slow instabilities:} \notag \\
\textrm{Unequal oscillation frequencies $\omega_{E_\nu} \neq \omega_{E_{\bar{\nu}}}$} \notag \\
\notag \\
\textbf{Fast instabilities:} \notag \\
\textrm{Unequal angular distributions $g_{\nu} \neq g_{\bar{\nu}}$} \notag \\
\notag \\
\textbf{Collisional instabilities:} \notag \\
\textrm{Unequal collisional rates $\Gamma \neq \bar{\Gamma}$} \notag \\
\notag
\end{gather}
Collective instabilities are driven by the neutrino--neutrino forward scattering potential, which is an integral of the neutrino--antineutrino flavor asymmetry. Thus collective instabilities often originate from asymmetries that cause neutrinos and antineutrinos to evolve differently in flavor space. Asymmetric collisional rates are indeed a reliable feature of compact-object environments. Accordingly, based on analytic estimates, collisional instabilities might appear in the neutrino decoupling regions of supernovae and mergers \cite{johns2021b}.

Ref.~\cite{johns2021b} established that collisional instabilities can appear in isotropic models. In such cases transient isotropization is not a concern. The isotropic calculations of that paper are self-consistent in the sense of Ref.~\cite{johns2022}: the initial conditions are steady states of the equations of motion without oscillation terms. It has also been observed, though, that FFC can kickstart collisional instability (see Fig.~3 in Ref.~\cite{johns2021b}). Anisotropy is a prerequisite for FFC, and so this particular observation \textit{does} face self-consistency concerns along the lines of those faced by scattering-enhanced FFC. Is FFC-hastened collisional instability also an artifact of modeling collisional, anisotropic flavor evolution using the homogeneous approximation?

In this paper we elaborate on the development of collisional instabilities in anisotropic, homogeneous models. We show that anisotropic collisional instability is possible without FFC accompanying it, that FFC is capable of expediting its development, and that in neither case is collisional instability a result of transient behavior.

To demonstrate this last point, we follow the method of Ref.~\cite{johns2022} and compare the evolution under different implementations of collisions. In one of these implementations, we allow collisions to act on the off-diagonal elements of the density matrices (the flavor coherences) but not on the diagonal elements (the number densities). This restriction prohibits collisions from reshaping the angular distributions. In real compact-object environments, angular distributions at a given location maintain quasi-steady anisotropy due to a balancing between advection and collisions. We are not even attempting to capture this aspect of the problem, which would necessarily require a spatially inhomogeneous model. Instead, we stick to a homogeneous model, artificially turn on or off collisional isotropization, and check whether the appearance of collisional instability hinges on this choice. Since isotropization is not self-consistently treated in a homogeneous model, it would be concerning if collisional instability, like scattering-enhanced FFC, turned out to critically depend on it.

Applying the method described above, we find that anisotropic collisional instabilities develop whether or not collisions are isotropizing. Moreover, they can be caused by neutral-current (NC) or charged-current (CC) scattering. These findings extend those of Ref.~\cite{johns2021b}, where, in the numerical results, collisional instability was driven principally by CC absorption and emission. 

Overall, through this work we gain a deeper understanding of how collisional instabilities evolve in homogeneous environments, which in turn provides clues as to how they might evolve in more realistic settings. In Sec.~\ref{sec:equations} we lay out the relevant equations of motion. In Sec.~\ref{sec:results1} we present the evolution of collisional instabilities in the absence of FFC, and in Sec.~\ref{sec:results2} we show the evolution in its presence. In Sec.~\ref{sec:discussion} we conclude.

\section{Equations of motion \label{sec:equations}}

We adopt a model that is homogeneous, axisymmetric, and monochromatic. We allow for vacuum oscillations, coherent neutrino--neutrino forward scattering, and incoherent scattering of a kind that we vary for comparative purposes.

The neutrino and antineutrino flavor density matrices $\rho_v$ and $\bar{\rho}_v$, for propagation angle $v = \cos\theta$, evolve in accord with the following equations of motion:
\begin{align}
&i \frac{d}{dt} \rho_v =  \left[ + H^\textrm{vac} + H^{\nu\nu}_v, \rho_v \right] + i C [\rho] \notag \\
&i \frac{d}{dt} \bar{\rho}_v = \left[ - H^\textrm{vac} + H^{\nu\nu}_v, \bar{\rho}_v \right] + i C [\bar{\rho}].
\end{align}
The vacuum Hamiltonian is
\begin{equation}
H_\textrm{vac} = \frac{\omega}{2}
\begin{pmatrix}
- \cos 2\theta & \sin 2\theta \\
\sin 2\theta & \cos 2\theta
\end{pmatrix} ,
\end{equation}
where $\omega = \delta m^2 / 2E_\nu$ is the vacuum oscillation frequency, $\delta m^2$ is the mass-squared splitting, $E_\nu$ is the neutrino energy, and $\theta$ is the mixing angle. The other part of the Hamiltonian is
\begin{equation}
H_v^{\nu\nu} = \sqrt{2} G_F \int d v' ( 1 - v v' ) ( \rho_{v'} - \bar{\rho}_{v'} ),
\end{equation}
arising from the forward scattering of neutrinos on each other.

Rather than implementing coherent neutrino--matter forward scattering explicitly, we replace the vacuum mixing angle by a small matter-suppressed value. This approximation is consistent with other recent literature on collisional effects in neutrino flavor evolution \cite{martin2021, shalgar2021, kato2021, sasaki2021, hansen2022, johns2021b, johns2022}. The rationale is that the potential generated by neutrino--matter forward scattering is independent of neutrino momentum and thus affects all flavor states in the same way. The chief effect is to suppress mixing of all neutrinos equally. Less trivial consequences of a matter background, like the multi-angle matter effect \cite{esteban2008}, are associated with the (momentum-dependent) advection term, which vanishes here due to the assumption of homogeneity.

The numerical values of the parameters, which are meant to represent the conditions in the neutrino decoupling region during the supernova accretion phase, are chosen to be nearly identical to those used in Ref.~\cite{johns2022}. The parameters are collected in Table~\ref{paramtab}. The neutrino and antineutrino scattering rates are taken to be $\Gamma = 50 \Gamma_{\nu n}$ and $\bar{\Gamma} = 5 \Gamma_{\nu n}$, respectively, where the neutrino--neutron scattering rate $\Gamma_{\nu n}$ is pulled from Ref.~\cite{johns2022}. To bring out the collisional instabilities more clearly in the numerical results, we use exaggeratedly large and asymmetric scattering rates. In the results that follow, we compare the normal and inverted mass hierarchies.

\begin{table*}
\centering
\begin{tabular}{|c|c|c|}
 \hline
 \multicolumn{3}{|c|}{Calculation parameters} \\
 \hline\hline
 ~~Variable~~ & ~~Meaning~~ & ~~Value~~ \\
 \hline
$\varrho$ & Density of the medium & $10^{12}$~g/cm$^3$ \\
$T$ & Temperature of the medium & 7 MeV \\
$\mu_e$ & Electron chemical potential & 20 MeV \\
$Y_e$ & Electron fraction & 0.13 \\
$E_\nu$ & Neutrino energy & 20 MeV \\
$\delta m^2$ & Mass-squared splitting & $2.4 \times 10^{-3}$~eV$^2$ \\
$\theta$ & Matter-suppressed mixing angle & $10^{-6}$ \\
$\omega$ & Vacuum oscillation frequency & 0.3 km$^{-1}$ \\
$n_{\nu_e} (0)$ & Initial number density of $\nu_e$ & $2.6 \times 10^{33}$~cm$^{-3}$ \\
$n_{\bar{\nu}_e} (0)$ & Initial number density of $\bar{\nu}_e$ & $2.5 \times 10^{33}$~cm$^{-3}$ \\
$n_{\nu_x} (0)$ & Initial number density of $\nu_x$ & $1.0 \times 10^{33}$~cm$^{-3}$ \\
$\mu | \mathbf{D}_0 (0) |$ & Neutrino--neutrino forward-scattering potential & $3 \times 10^5$ km$^{-1}$ \\ 
$\Gamma$ & Neutrino scattering rate & 26 km$^{-1}$ \\ 
$\bar{\Gamma}$ & Antineutrino scattering rate & 2.6 km$^{-1}$ \\ 
 \hline
\end{tabular}
\caption{Parameters adopted for the numerical calculations. The number density of $\bar{\nu}_x$ is taken to be the same as that of $\nu_x$. The strength of the initial-time neutrino--neutrino forward-scattering potential is presented as $\mu | \mathbf{D}_0 (0)|$, $\mu$ is the usual self-coupling parameter and $\mathbf{D}_0 (0)$ is the the $l=0$ difference vector at $t = 0$ [Eq.~\eqref{eq:D0}]. Because of the initial symmetry between $\nu_x$ and $\bar{\nu}_x$, the forward-scattering potential is equal to $\sqrt{2} G_F [ n_{\nu_e} (0) - n_{\bar{\nu}_e} (0) ]$. See the main text and Fig.~\ref{ang_distrs} for specification of the angular distributions. \label{paramtab}}
\end{table*}

\begin{figure}
\begin{subfigure}{
\centering
\includegraphics[width=.42\textwidth]{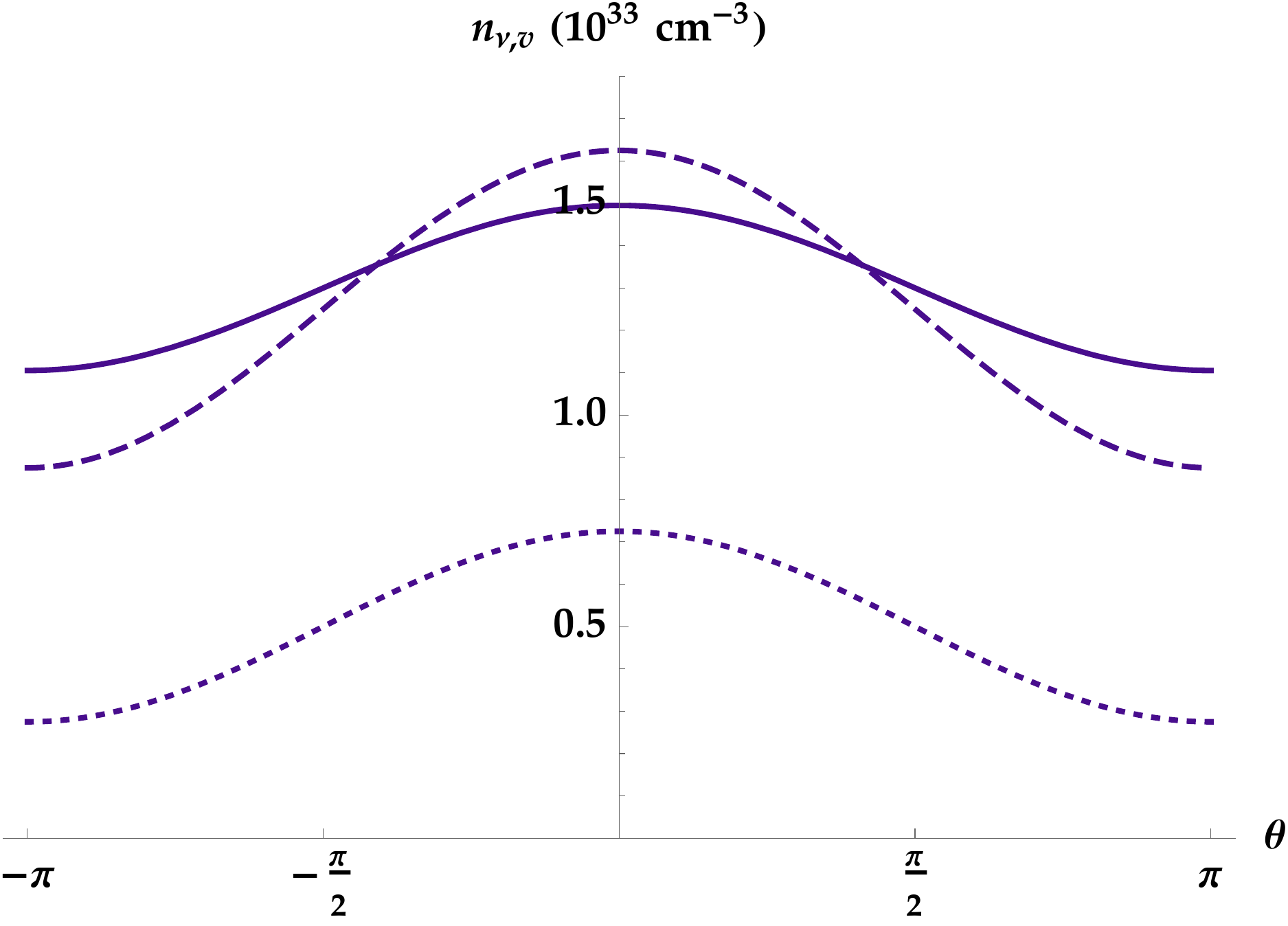}
}
\end{subfigure}

\begin{subfigure}{
\centering
\includegraphics[width=.42\textwidth]{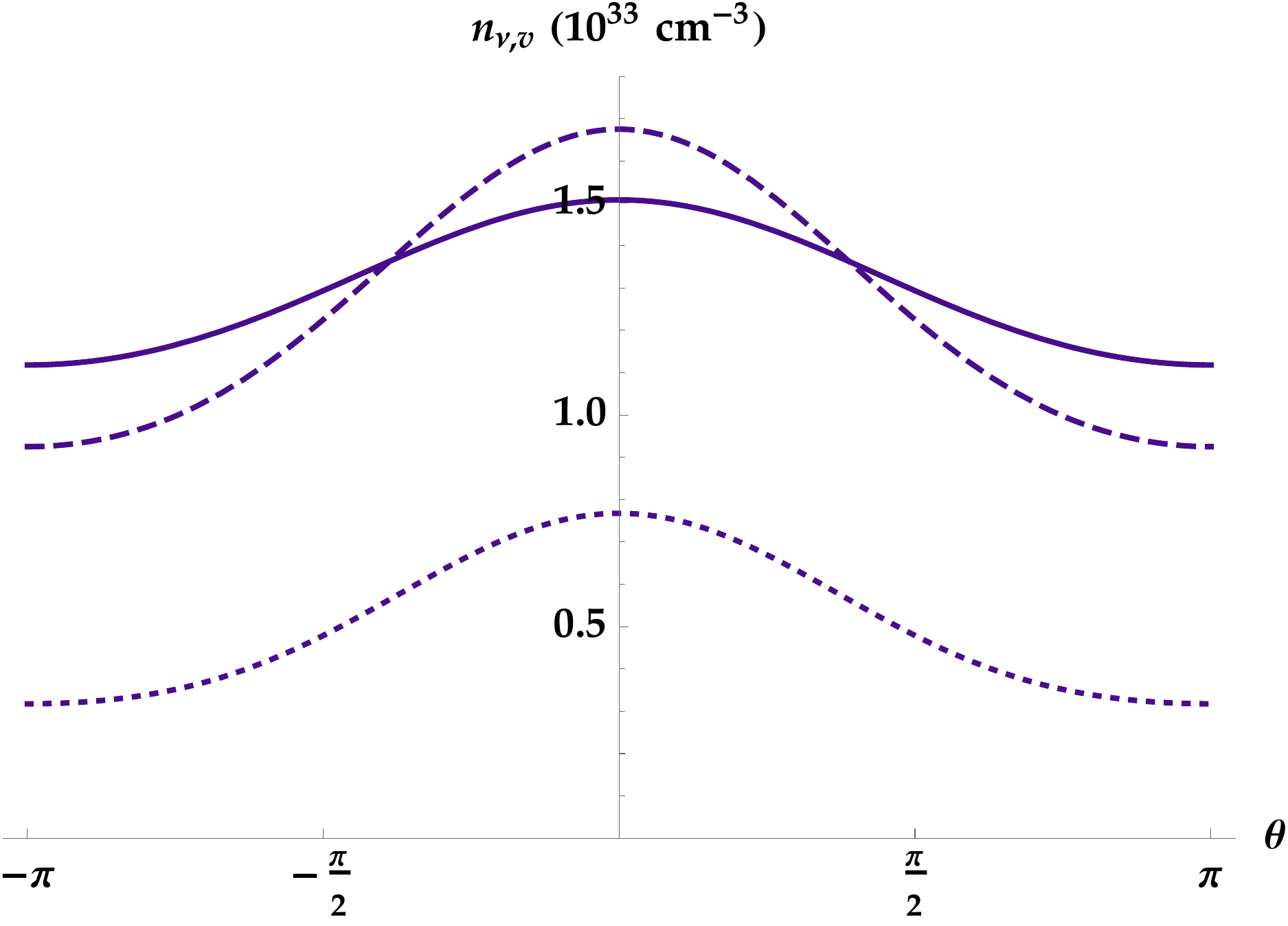}
}
\end{subfigure}
\caption{Neutrino angular distributions used in this study: $n_{\nu_e, v}$ (solid), $n_{\bar{\nu}_e, v}$ (dashed), $n_{\nu_x, v} = n_{\bar{\nu}_x, v}$ (dotted), where $v = \cos\theta$ and $\theta$ is the neutrino propagation angle. The lower panel shows the same distributions used in Ref.~\cite{johns2022}. The distributions in the upper panel are identical but with the $l = 2$ Legendre moments zeroed out. Although the two panels look very similar and both contain angular crossings, only the lower panel is unstable to fast flavor conversion \cite{johns2020}. \label{ang_distrs}}
\end{figure}

In order to observe collisionally unstable evolution both with and without FFC, we compare two sets of angular distributions. In the case where the system is unstable to collisional and fast instabilities, the angular distributions are taken to be
\begin{equation}
n_{\nu_\alpha, v} = \frac{1}{2} n_{\nu_\alpha, 0} L_0 (v) + \frac{3}{2} n_{\nu_\alpha, 1} L_1 (v) + \frac{5}{2} n_{\nu_\alpha, 2} L_2 (v), \label{eq:ang_ffc}
\end{equation}
with $n_{\nu_\alpha, l}$ the $l$th Legendre moment of the angular distribution of flavor $\alpha$, $L_l (v)$ the $l$th Legendre polynomial, and $n_{\nu_\alpha, 0} = n_{\nu_\alpha}$ as specified above. For $\nu_e$ we use the number-density moments
\begin{align}
&n_{\nu_e, 1} = 0.050 n_{\nu_e} \notag \\
&n_{\nu_e, 2} = 0.002 n_{\nu_e},
\end{align}
for $\bar{\nu}_e$ we use
\begin{align}
&n_{\bar{\nu}_e, 1} = 0.100 n_{\bar{\nu}_e} \notag \\
&n_{\bar{\nu}_e, 2} = 0.008 n_{\bar{\nu}_e}, 
\end{align}
and for $\nu_x$ we use
\begin{align}
&n_{\nu_x, 1} = 0.150 n_{\nu_x} \notag \\
&n_{\nu_x, 2} = 0.017 n_{\nu_x}.
\end{align}
The $\bar{\nu}_x$ distribution is the same as that of $\nu_x$. All higher moments are set to zero. In the other case, where the system is stable to fast oscillations, we use
\begin{equation}
n_{\nu_\alpha, v} = \frac{1}{2} n_{\nu_\alpha, 0} L_0 (v) + \frac{3}{2} n_{\nu_\alpha, 1} L_1 (v), \label{eq:ang_noffc}
\end{equation}
which is equivalent to dropping the $l=2$ moments, keeping the other moments the same. The distributions are plotted in Fig.~\ref{ang_distrs}. Although zeroing out the $l=2$ moments only changes the distributions slightly, and does not change the presence of an angular crossing, in a homogeneous setting it makes the difference between being fast-stable and fast-unstable \cite{johns2020}. This is related to the fact that homogeneity isolates the $K = 0$ Fourier mode of $\rho$---all $K \neq 0$ being forced to vanish by symmetry---and limits the range of fast instabilities that can be expressed by the system.

Following Ref.~\cite{johns2022}, we consider three different collisional prescriptions, implementing them one at a time and comparing the results. The first is
\begin{equation}
\textbf{\textrm{NC (isotropizing):}} ~~~ i\mathcal{C} =  \Gamma \left( \rho_\textrm{NC} - \rho \right),
\end{equation}
where
\begin{equation}
\rho_\textrm{NC} = \langle \rho \rangle = 
\begin{pmatrix}
\langle \rho \rangle_{ee} & \langle \rho \rangle_{ex} \\
\langle \rho \rangle_{xe} & \langle \rho \rangle_{xx}
\end{pmatrix}
\end{equation}
and
\begin{equation}
\langle \rho \rangle = \frac{1}{2} \int_{-1}^{+1} dv \rho_v.
\end{equation}
This is a simplified way of describing, for example, neutrino--nucleon scattering. The assumption is that the scattering rate is independent of the ingoing and outgoing neutrino momenta. This is the collisional form adopted in, for example, Ref.~\cite{shalgar2021}. Although a more realistic implementation of NC scattering would involve a momentum-dependent scattering kernel, the differences are not qualitatively significant \cite{martin2021}.

The second prescription is
\begin{equation}
\textbf{\textrm{CC (isotropizing):}} ~~~ i\mathcal{C} =  \Gamma \left( \rho_\textrm{CC} - \rho \right),
\end{equation}
where
\begin{equation}
\rho_\textrm{CC} = 
\begin{pmatrix}
\langle \rho \rangle_{ee} & 0 \\
0 & \langle \rho \rangle_{xx}
\end{pmatrix}. \label{eq:ccisoeq}
\end{equation}
This collisional form approximates the effect of CC neutrino--electron scattering. It is identical to isotropizing NC scattering except for the vanishing of flavor coherence in the collisional equilibrium state given by Eq.~\eqref{eq:ccisoeq}. Unlike NC scattering, CC scattering is sensitive to the flavor state of the scattering neutrino and therefore causes flavor decoherence.

The third and final prescription is
\begin{equation}
\textbf{\textrm{CC (non-isotropizing):}} ~~~ i\mathcal{C} =  - \Gamma \rho_T,
\end{equation}
where
\begin{equation}
\rho_T = \begin{pmatrix}
0 & \rho_{ex} \\
\rho_{xe} & 0
\end{pmatrix}.
\end{equation}
This form is not meant to describe any particular process, but rather to isolate the effect of collisional decoherence on the flavor evolution. As is evident from the equations above, non-isotropizing CC scattering has no direct influence on the diagonal (\textit{i.e.}, number-density) elements of the density matrices.

By comparing \textit{NC (isotropizing)} and \textit{CC (isotropizing)}, we establish whether the charged- or neutral-current nature of the interaction is important for the dynamics we observe, including the development of the collisional instability and the late-time flavor-conversion outcomes. In comparing to \textit{CC (non-isotropizing)}, we are additionally checking whether the observed dynamics is a product of the (astrophysically unrealistic) isotropization that inevitably occurs in a homogeneous model with isotropizing collisional processes. Because \textit{CC (non-isotropizing)} acts only to damp the off-diagonal elements of the density matrices, if those elements already vanish, then the flavor distributions are at an equilibrium of this collisional term. In all of the calculations we present, the initial state has no flavor coherence, and so in the \textit{CC (non-isotropizing)} calculations, the system begins at an equilibrium of the collisional part of the equations of motion. With the other two implementations, the system begins out of equilibrium due to collisions alone.

The value of this three-way comparison test is described in more detail in Ref.~\cite{johns2022}. There it was found that scattering-enhanced FFC does not survive in calculations implementing \textit{CC (non-isotropizing)}, a finding that suggests that this particular phenomenon is an artifact of isotropization of the angular distributions---a feature that is not expected in real astrophysical settings. Below we report that collisional instabilities appear in all three collisional implementations. This finding on its own does not necessarily imply that collisional instabilities develop in real sites, but it does exclude the possibility that collisional instabilities are modeling artifacts in the same way that scattering-enhanced FFC is.

All calculations in this paper were run with 4000 angular moments, enough to achieve convergence.

\section{Collisional instability \textit{without} FFC \label{sec:results1}}

\begin{figure*}
\centering
\begin{subfigure}{
\centering
\includegraphics[width=.310\textwidth]{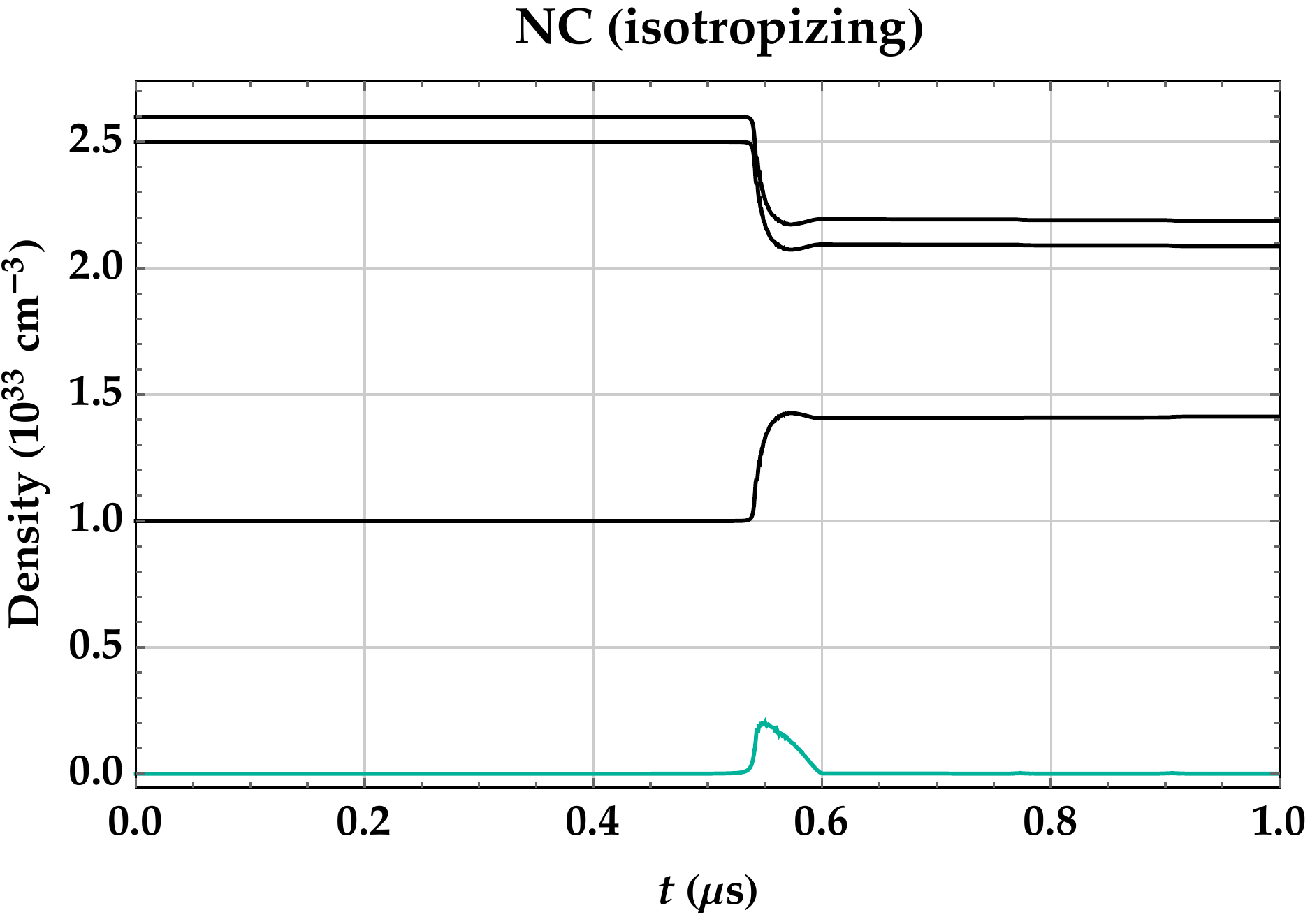}
}
\end{subfigure}
\begin{subfigure}{
\centering
\includegraphics[width=.310\textwidth]{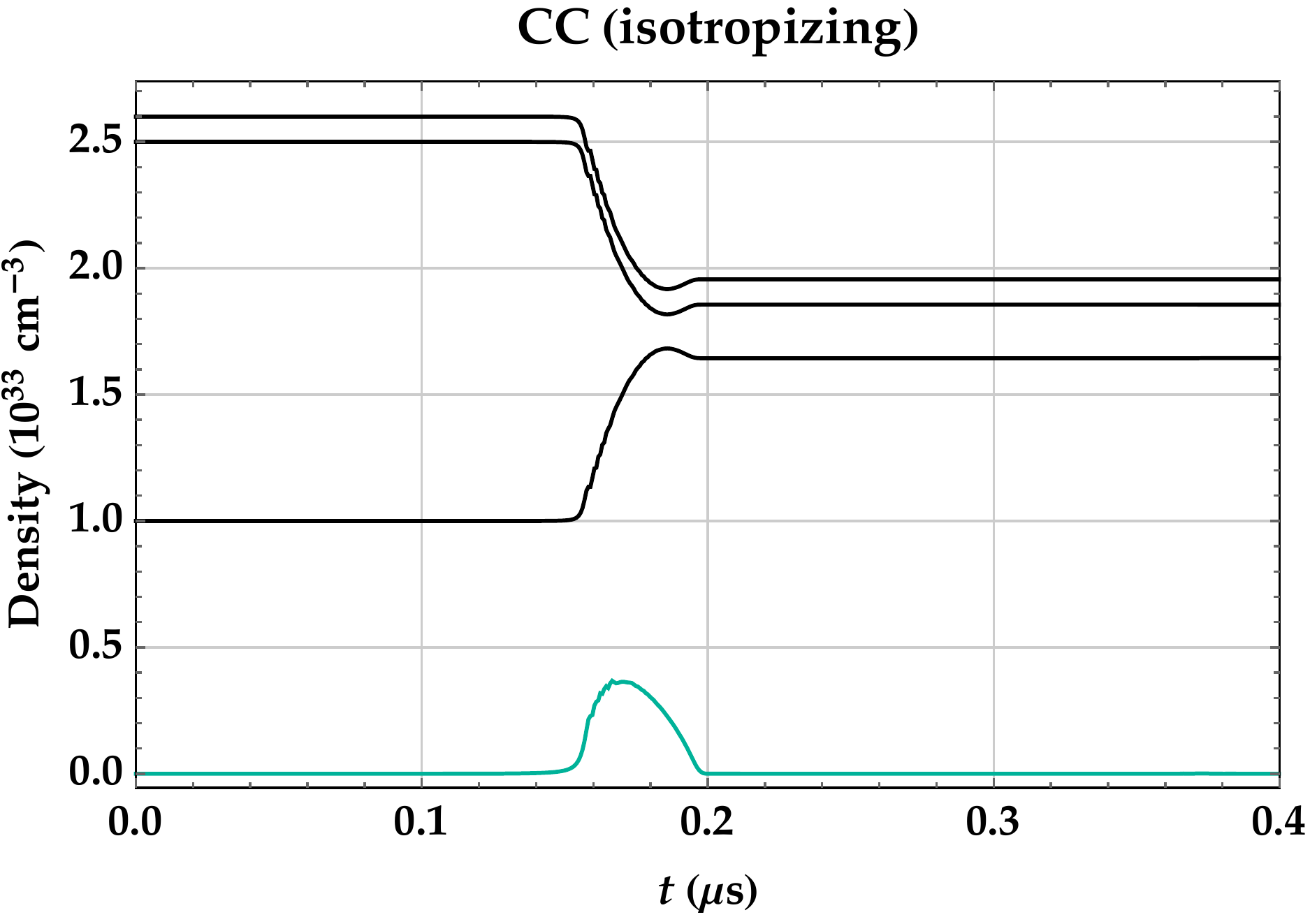}
}
\end{subfigure}
\begin{subfigure}{
\centering
\includegraphics[width=.310\textwidth]{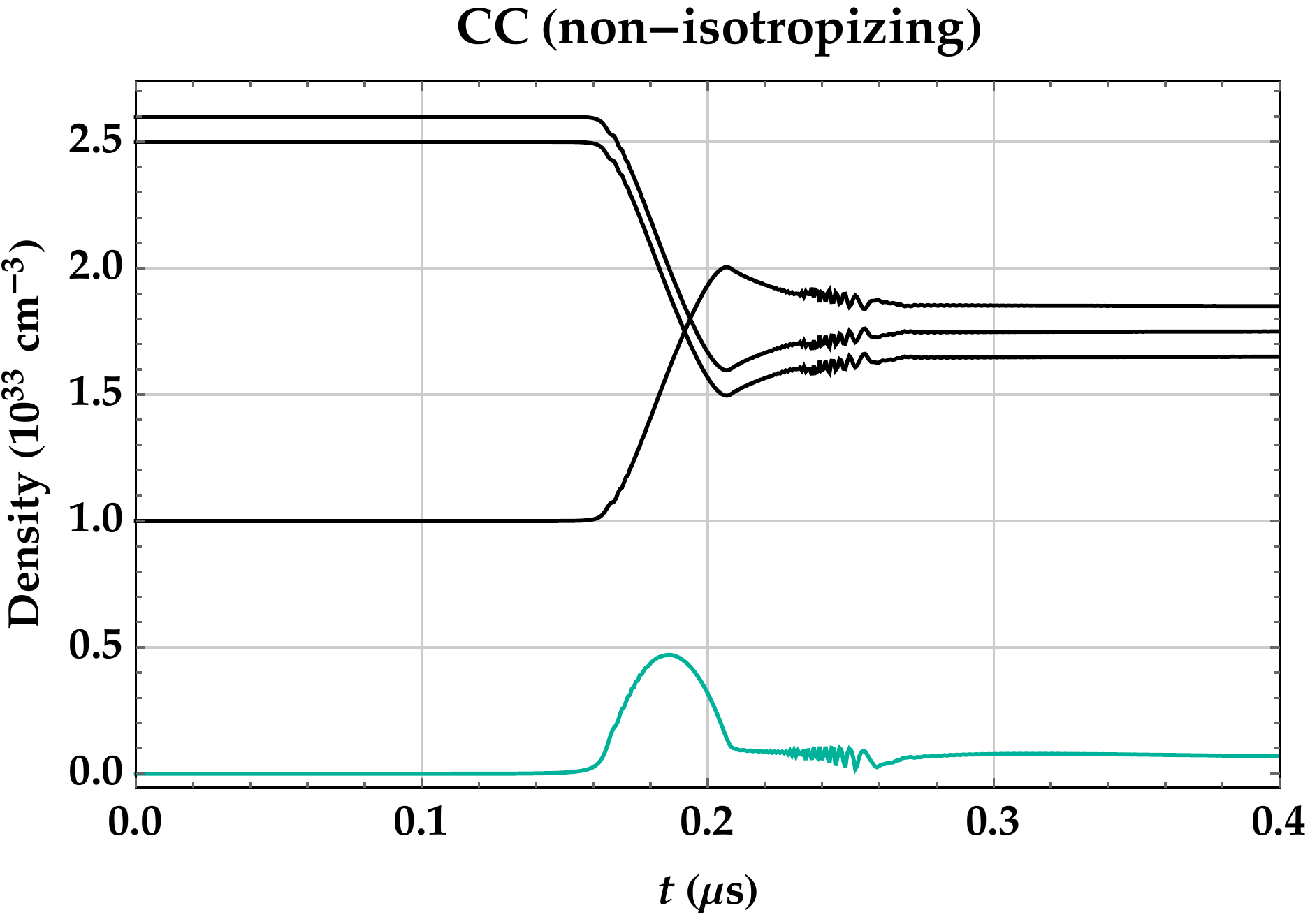}
}
\end{subfigure}

\begin{subfigure}{
\centering
\includegraphics[width=.310\textwidth]{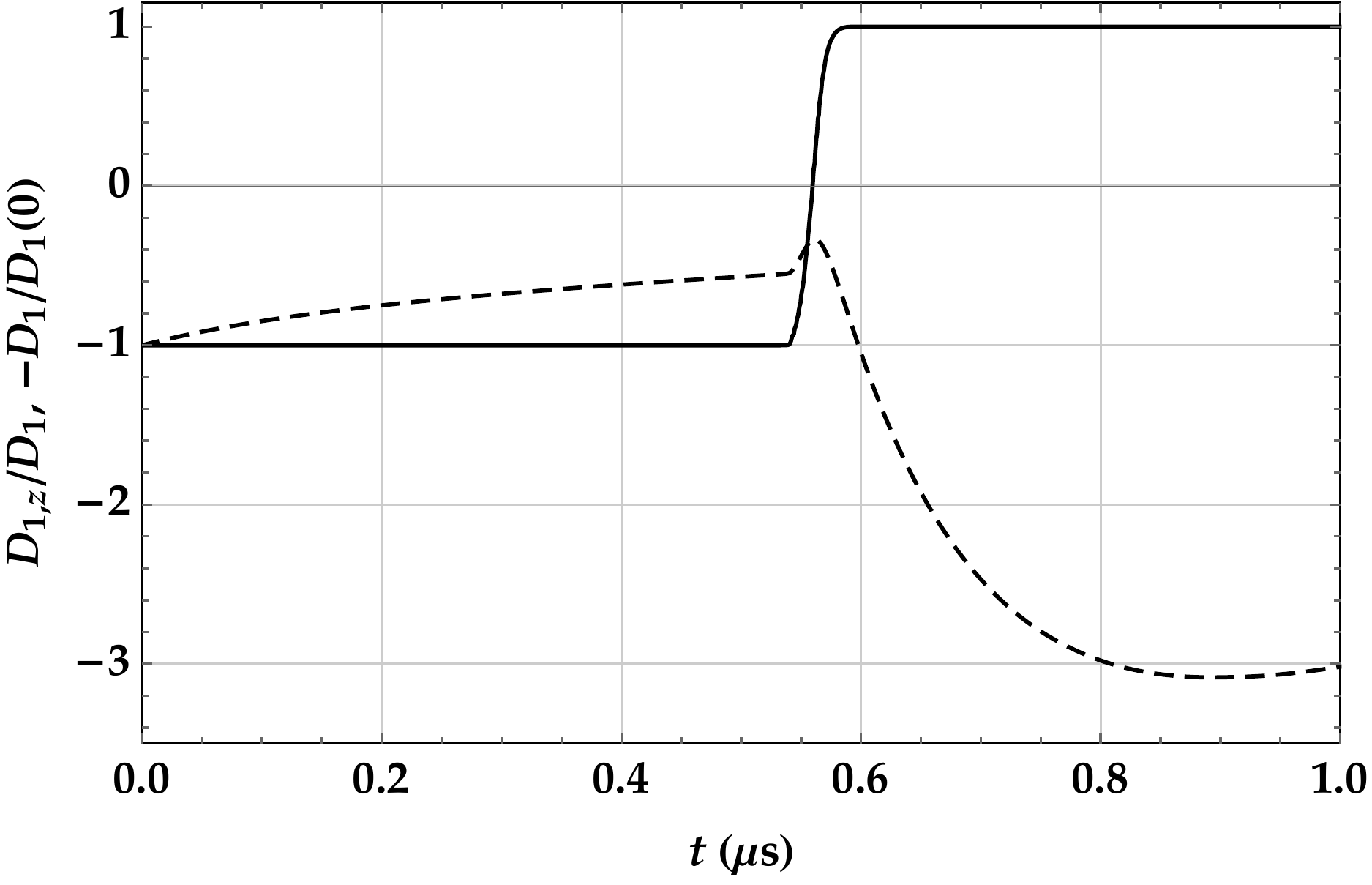}
}
\end{subfigure}
\begin{subfigure}{
\centering
\includegraphics[width=.310\textwidth]{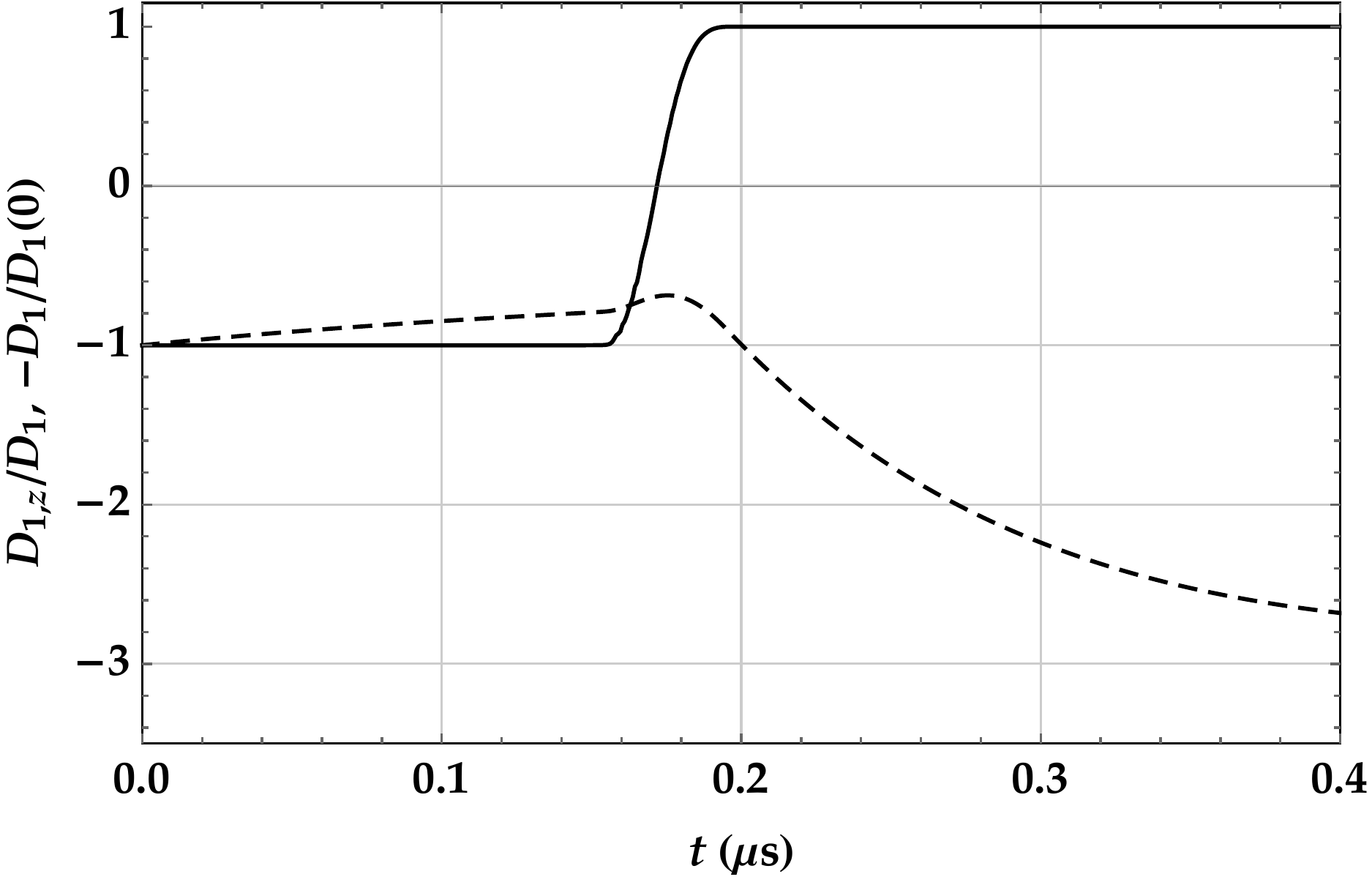}
}
\end{subfigure}
\begin{subfigure}{
\centering
\includegraphics[width=.310\textwidth]{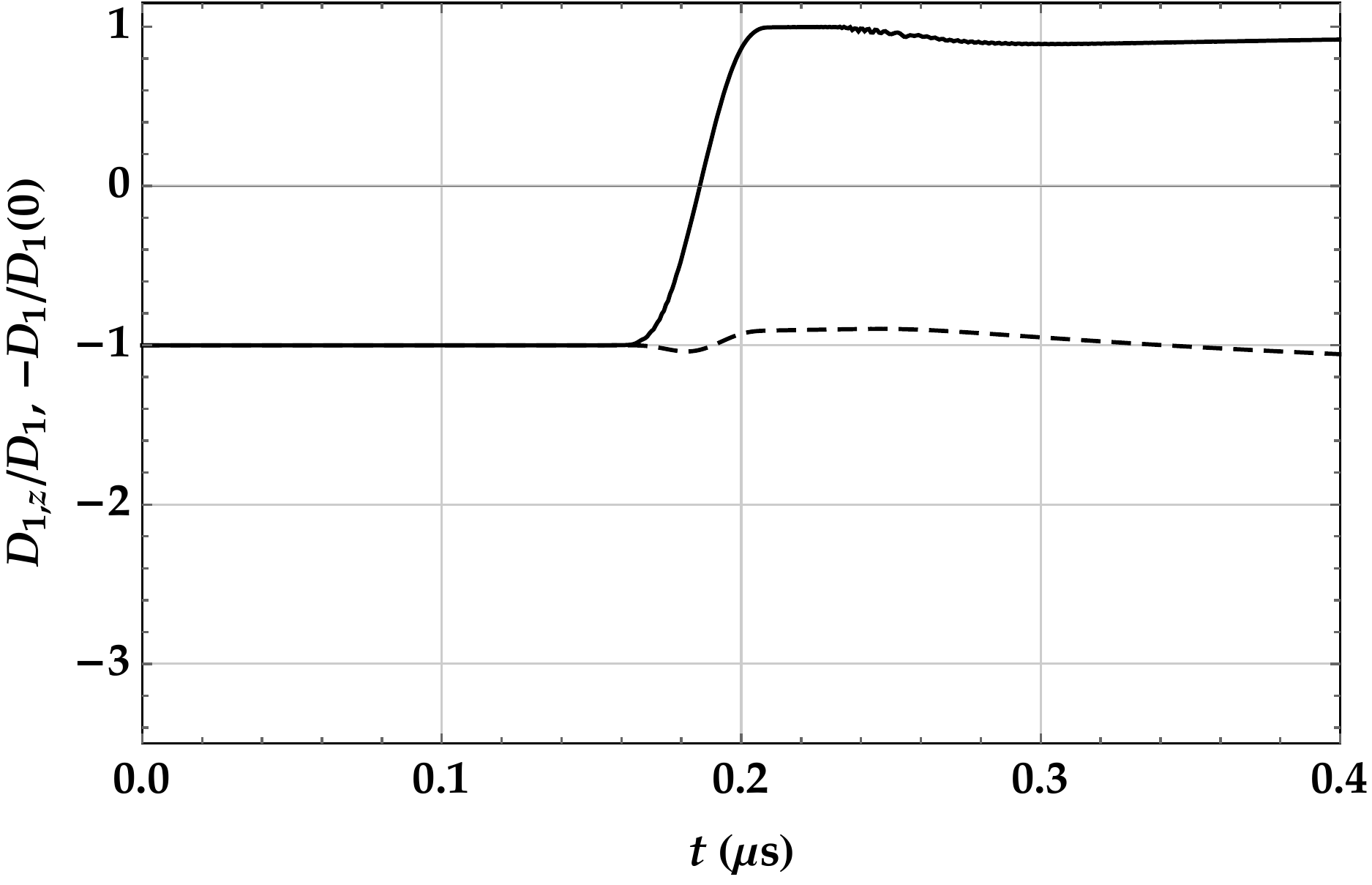}
}
\end{subfigure}

\begin{subfigure}{
\centering
\includegraphics[width=.310\textwidth]{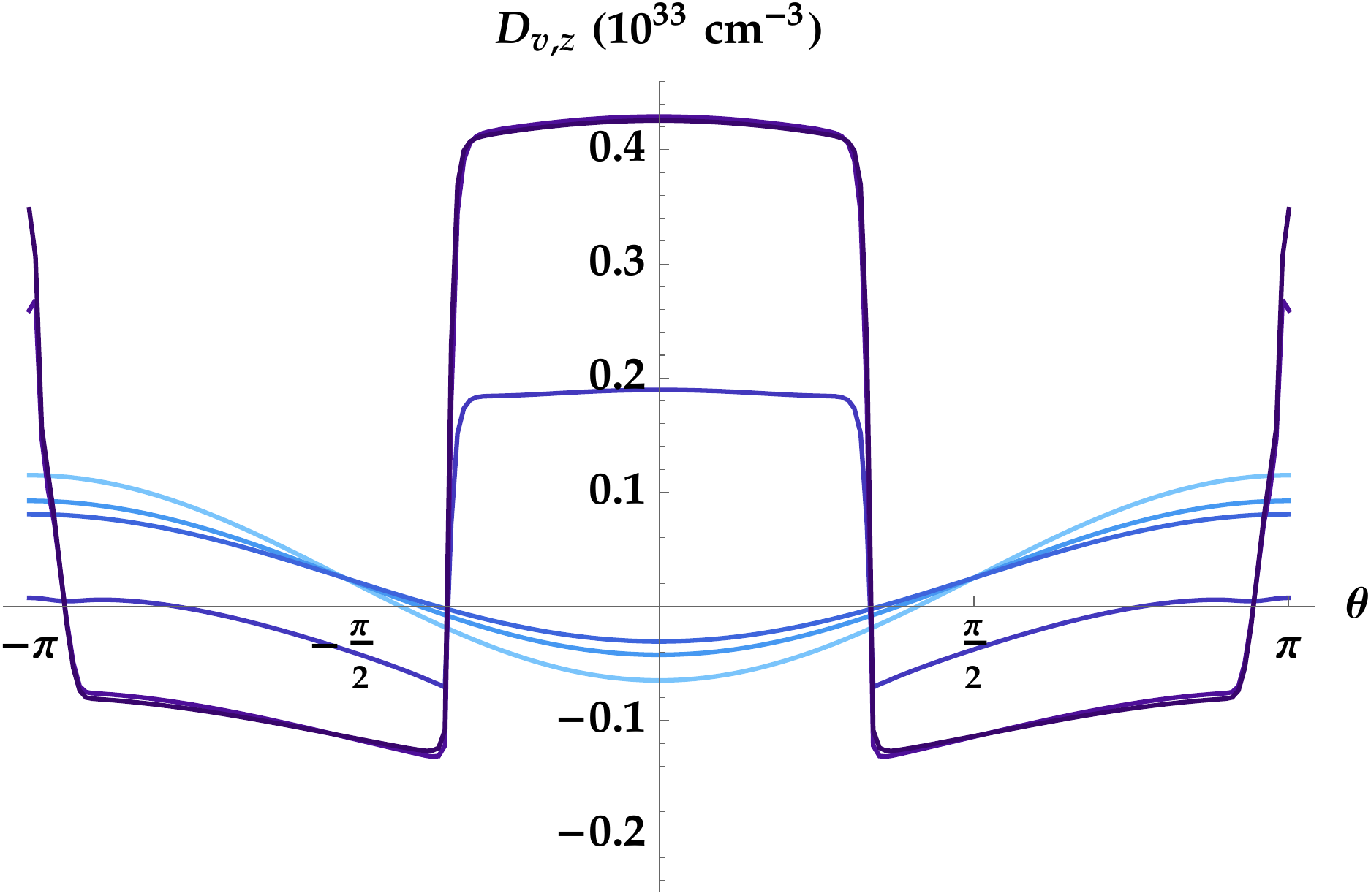}
}
\end{subfigure}
\begin{subfigure}{
\centering
\includegraphics[width=.310\textwidth]{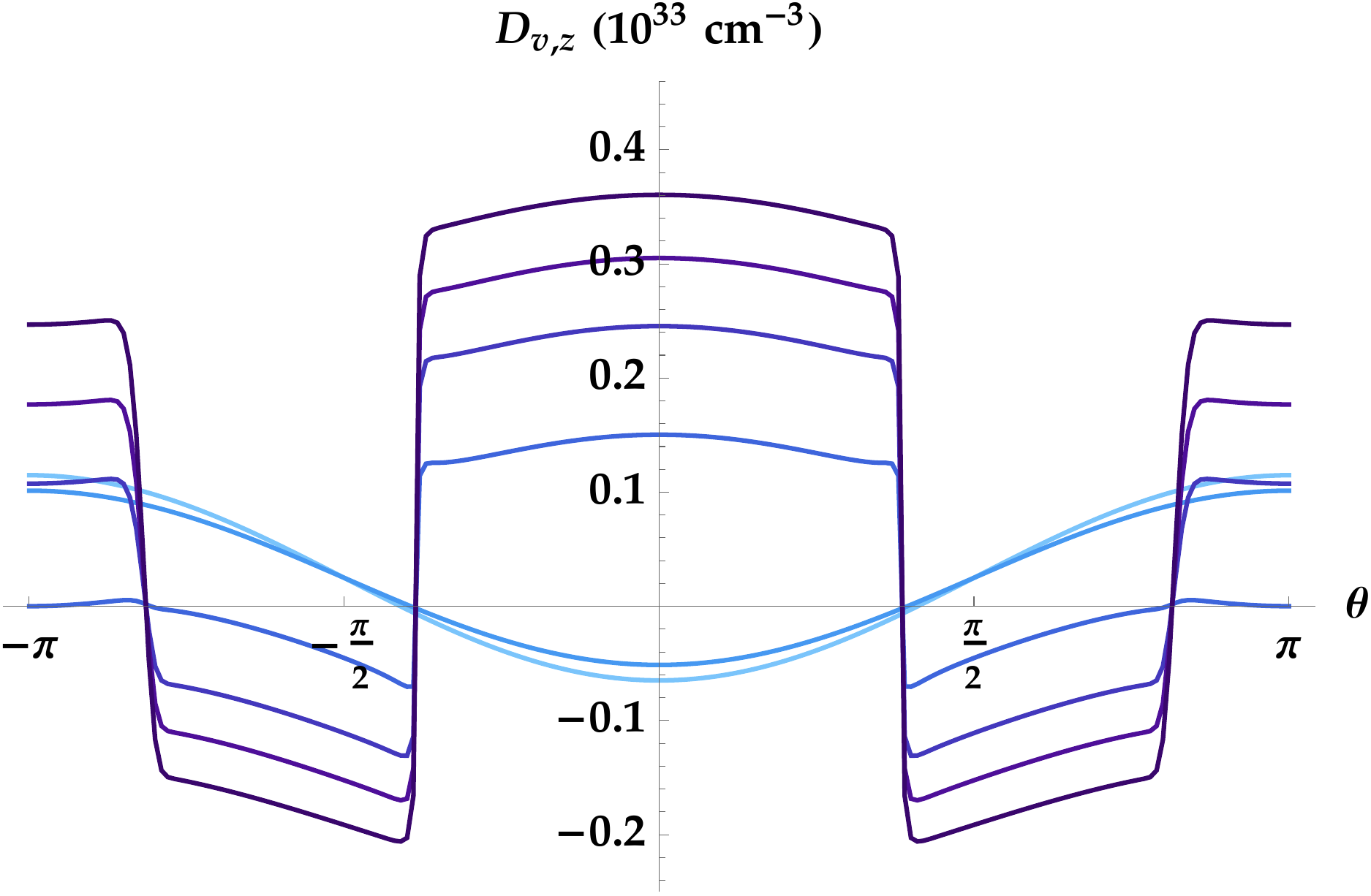}
}
\end{subfigure}
\begin{subfigure}{
\centering
\includegraphics[width=.310\textwidth]{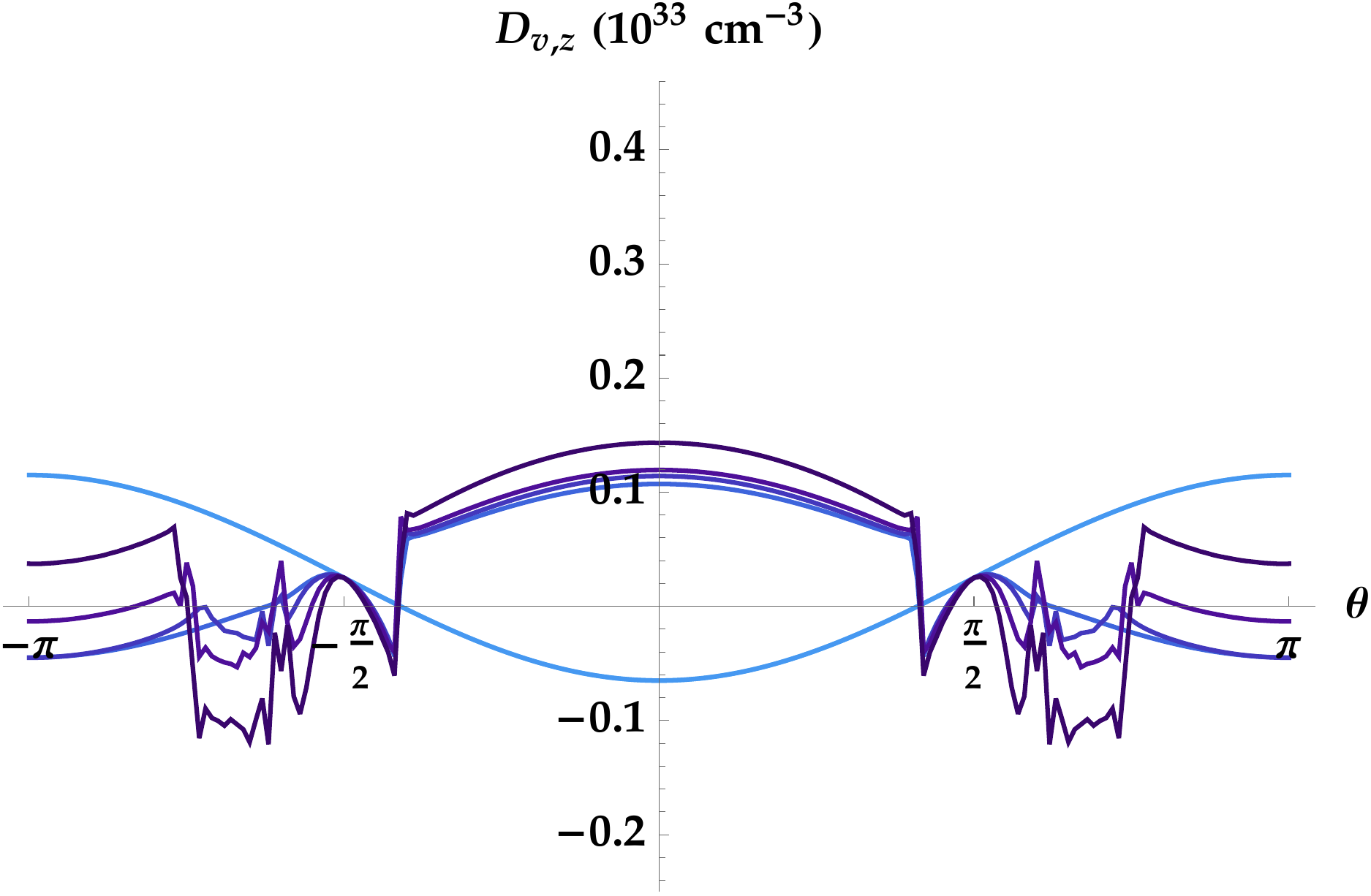}
}
\end{subfigure}

\begin{subfigure}{
\centering
\includegraphics[width=.310\textwidth]{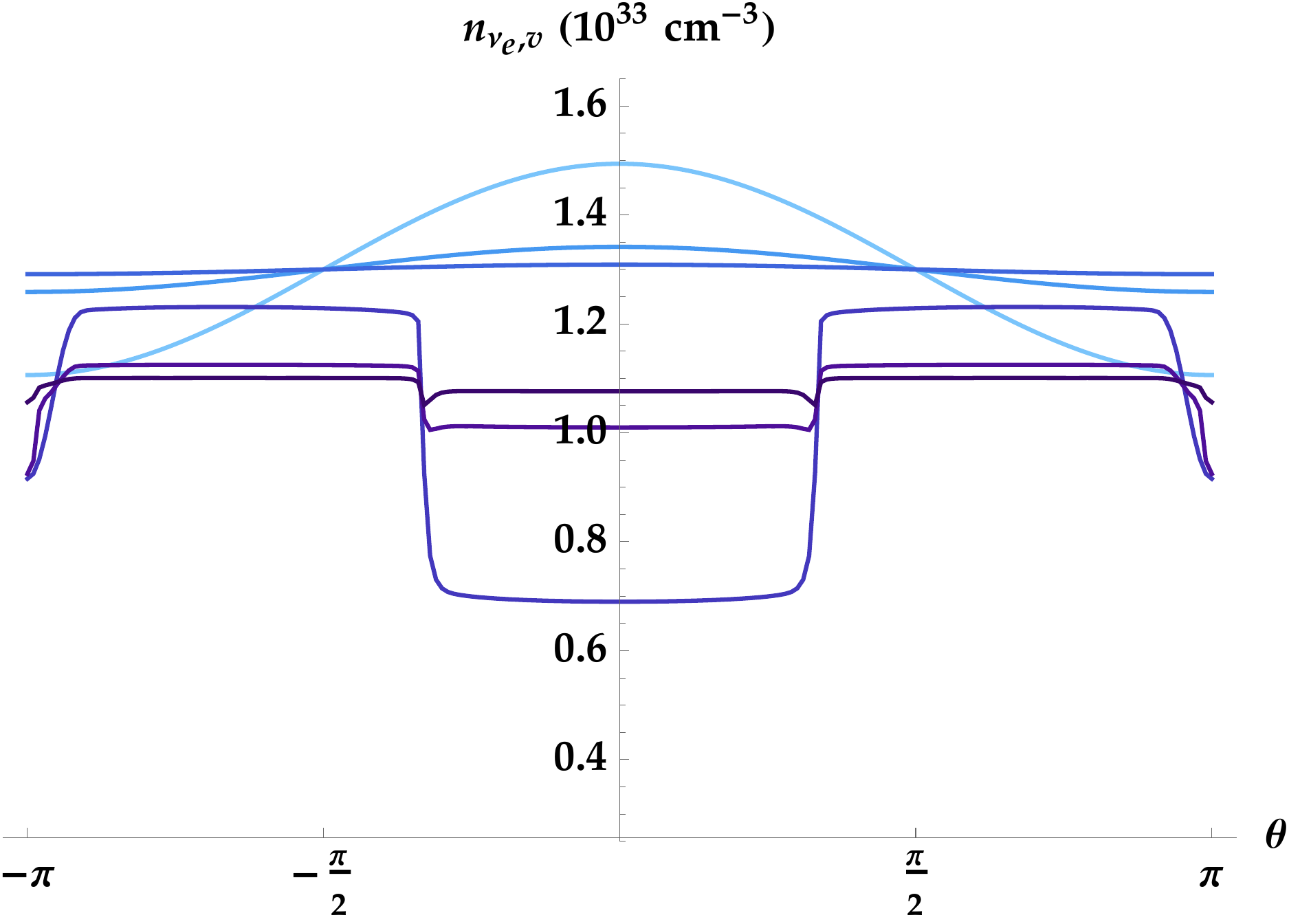}
}
\end{subfigure}
\begin{subfigure}{
\centering
\includegraphics[width=.310\textwidth]{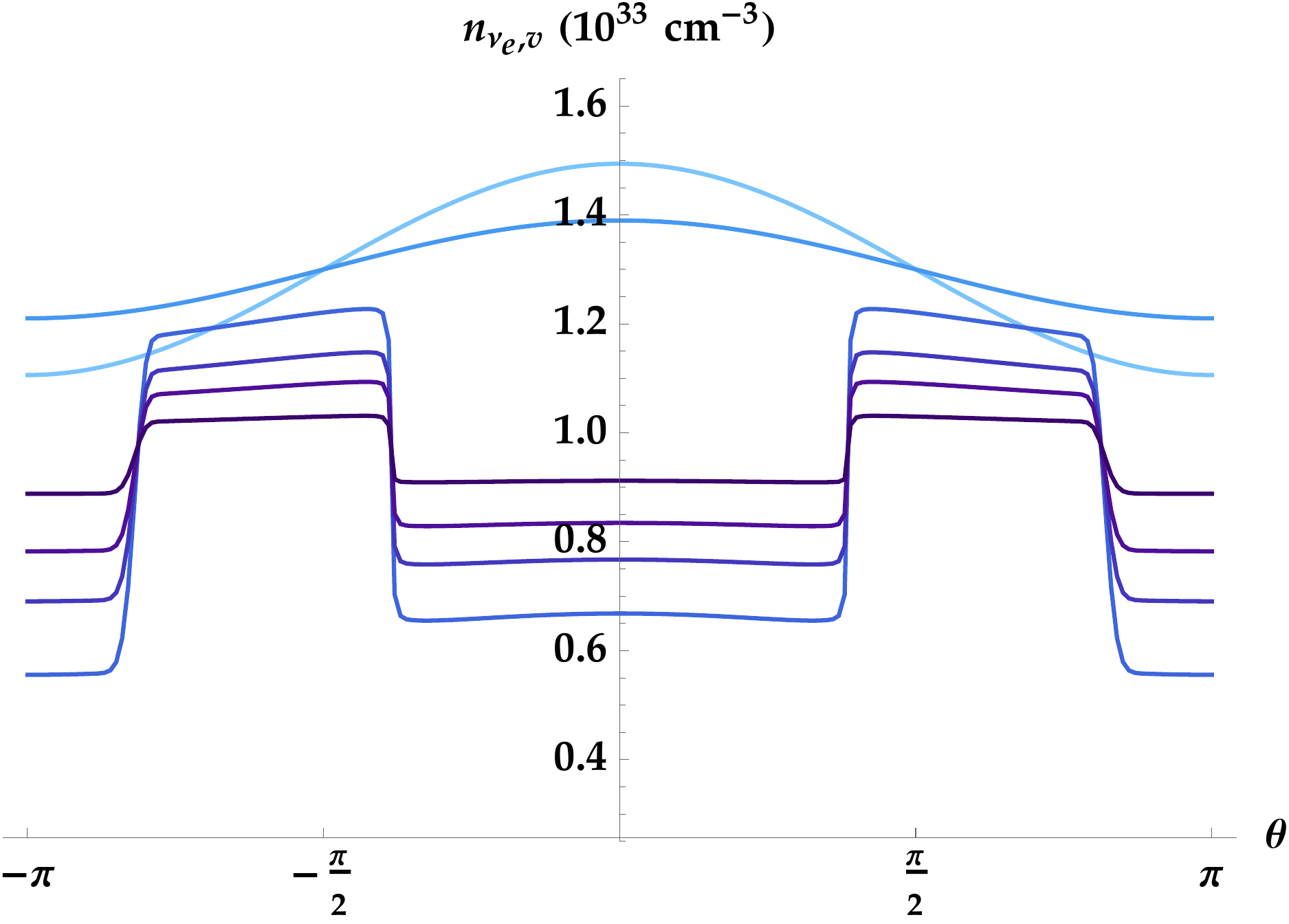}
}
\end{subfigure}
\begin{subfigure}{
\centering
\includegraphics[width=.310\textwidth]{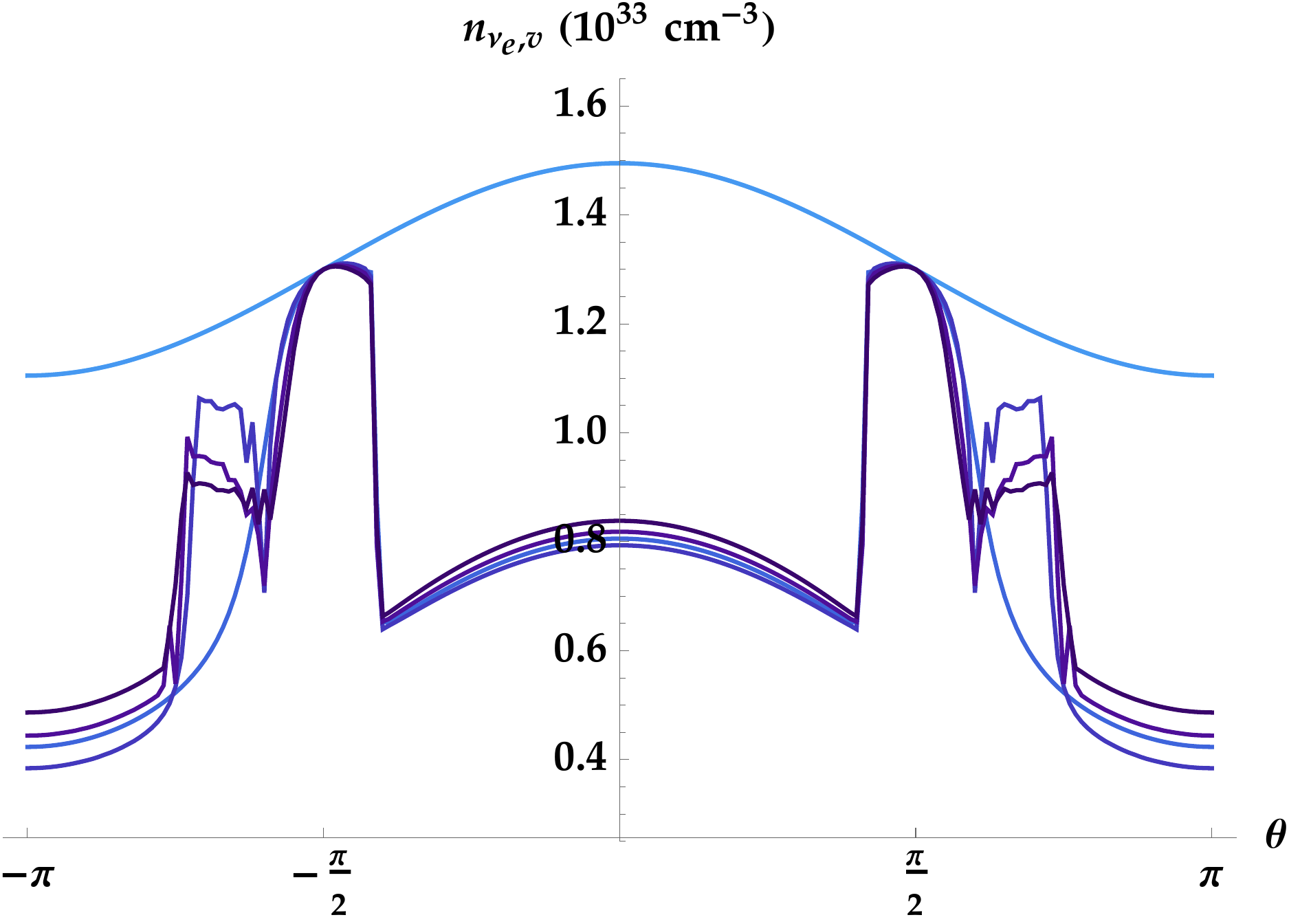}
}
\end{subfigure}
\caption{Collisionally unstable flavor evolution \textit{without} fast instability [angular distributions are given by Eq.~\eqref{eq:ang_noffc}] in the \textit{normal} mass hierarchy. \textit{First row:} Evolution of the number densities $n_{\nu_e}$ (top black curve), $n_{\bar{\nu}_e}$ (middle), $n_{\nu_x}$ (bottom), and neutrino coherence density $| \mathbf{P}_{0,T} | / 2$ (teal), where $T$ signifies the part orthogonal to the flavor ($\mathbf{z}$) axis. \textit{Second:} Evolution of $D_{1,z} / D_1$ (solid) and $-D_1 / D_1 (0)$ (dashed). \textit{Third:} Snapshots of $D_{v,z}$, where $v = \cos \theta$, color-coded with the earliest snapshot time in the lightest shade. \textit{Fourth:} Snapshots of $n_{\nu_e, v}$ at the same times. The columns correspond to calculations with isotropizing NC scattering (\textit{left}), isotropizing CC scattering (\textit{middle}), and non-isotropizing CC scattering (\textit{right}), as defined in Sec.~\ref{sec:equations}. Note the different scales of the NC and CC cases. For NC scattering, the displayed snapshot times are $t = 0$, $0.20$, $0.40$, $0.60$, $0.80$, and $1.00~\mu$s. For CC scattering (both isotropizing and non-isotropizing), they are $t = 0$, $0.10$, $0.20$, $0.25$, $0.30$, and $0.40~\mu$s. All three implementations of scattering exhibit a collisional instability involving the vector $\mathbf{D}_1$ inverting its orientation. See the text for further details.}  
\label{noffc_nh}
\end{figure*}

\begin{figure*}
\centering
\begin{subfigure}{
\centering
\includegraphics[width=.310\textwidth]{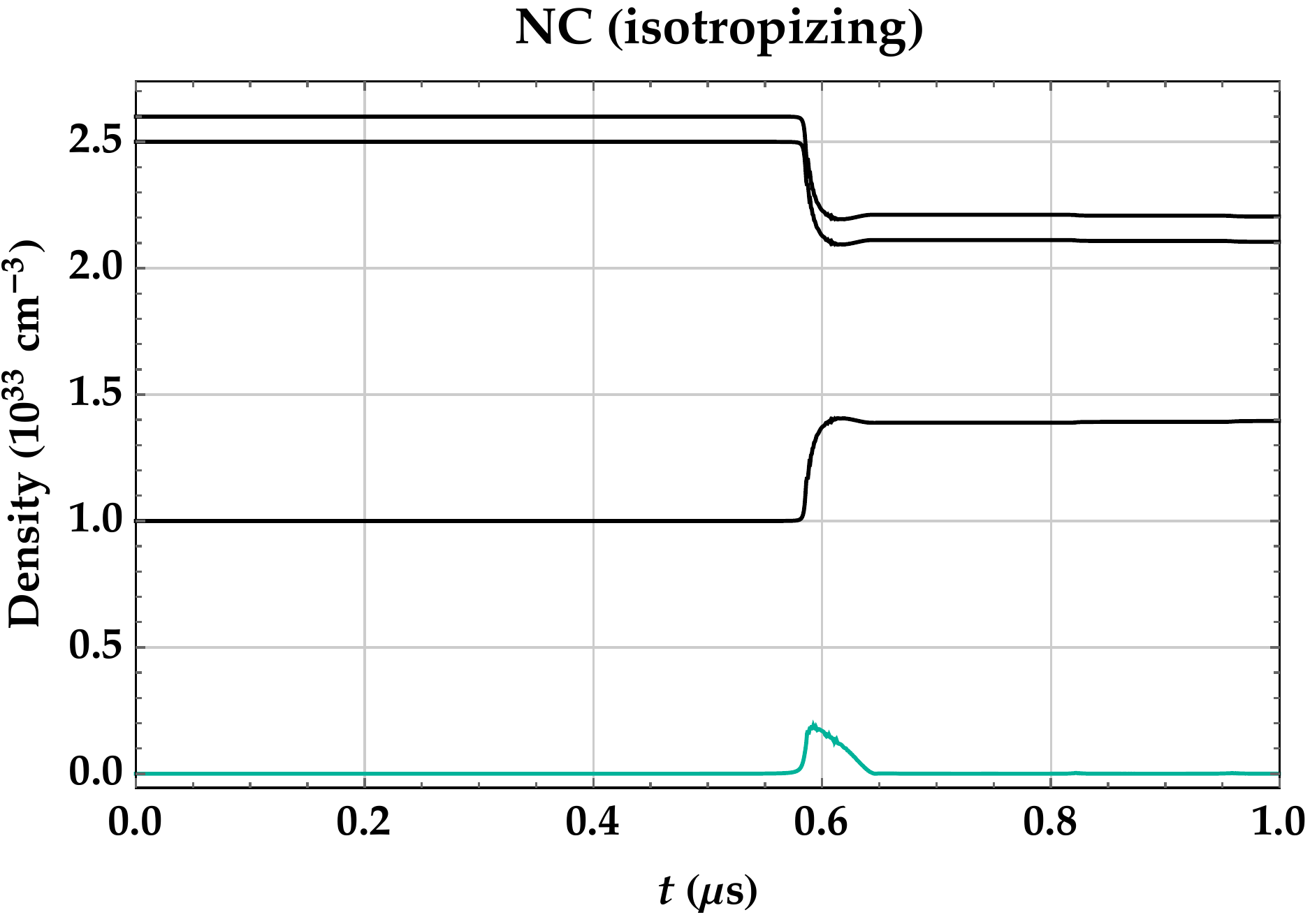}
}
\end{subfigure}
\begin{subfigure}{
\centering
\includegraphics[width=.310\textwidth]{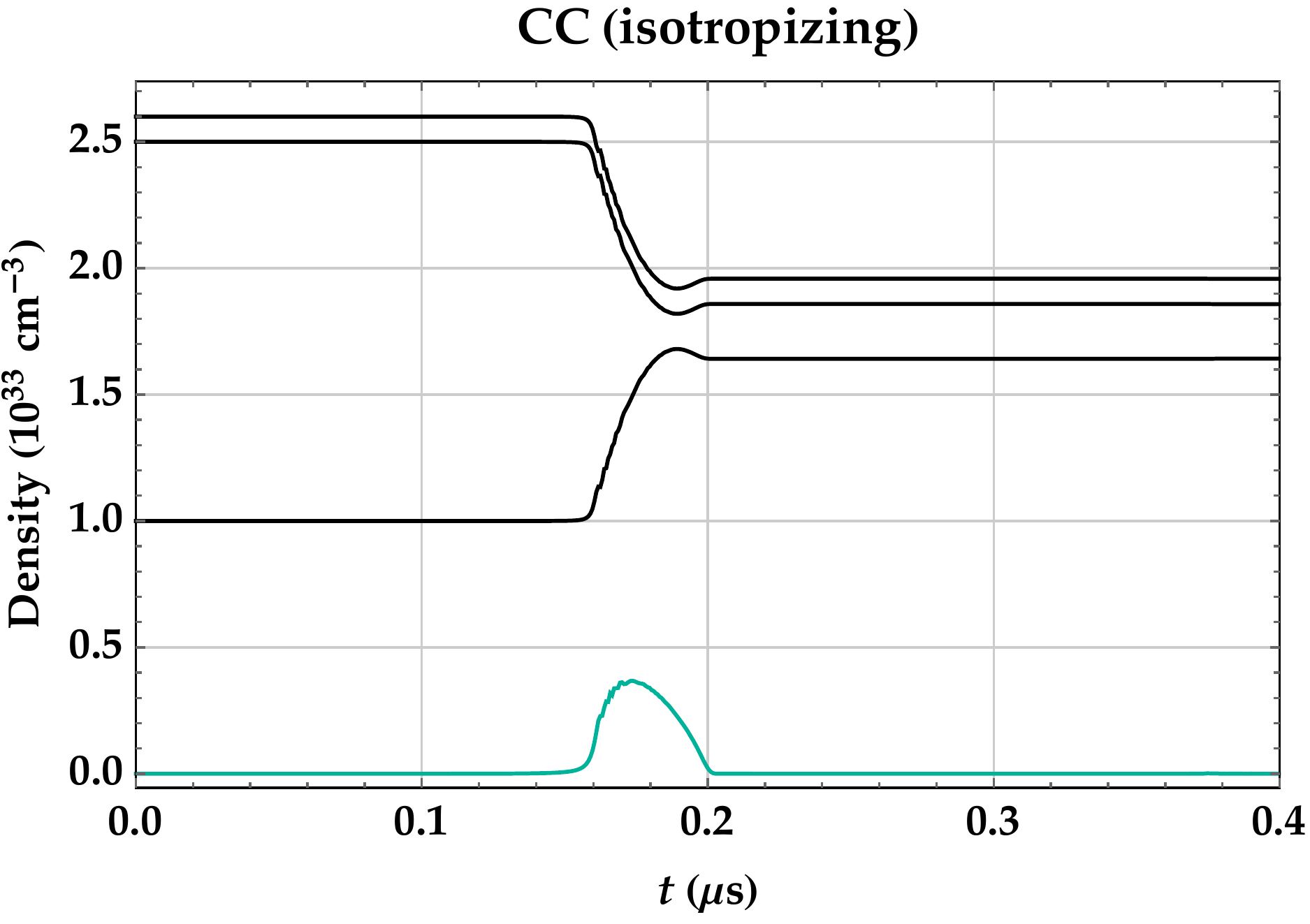}
}
\end{subfigure}
\begin{subfigure}{
\centering
\includegraphics[width=.310\textwidth]{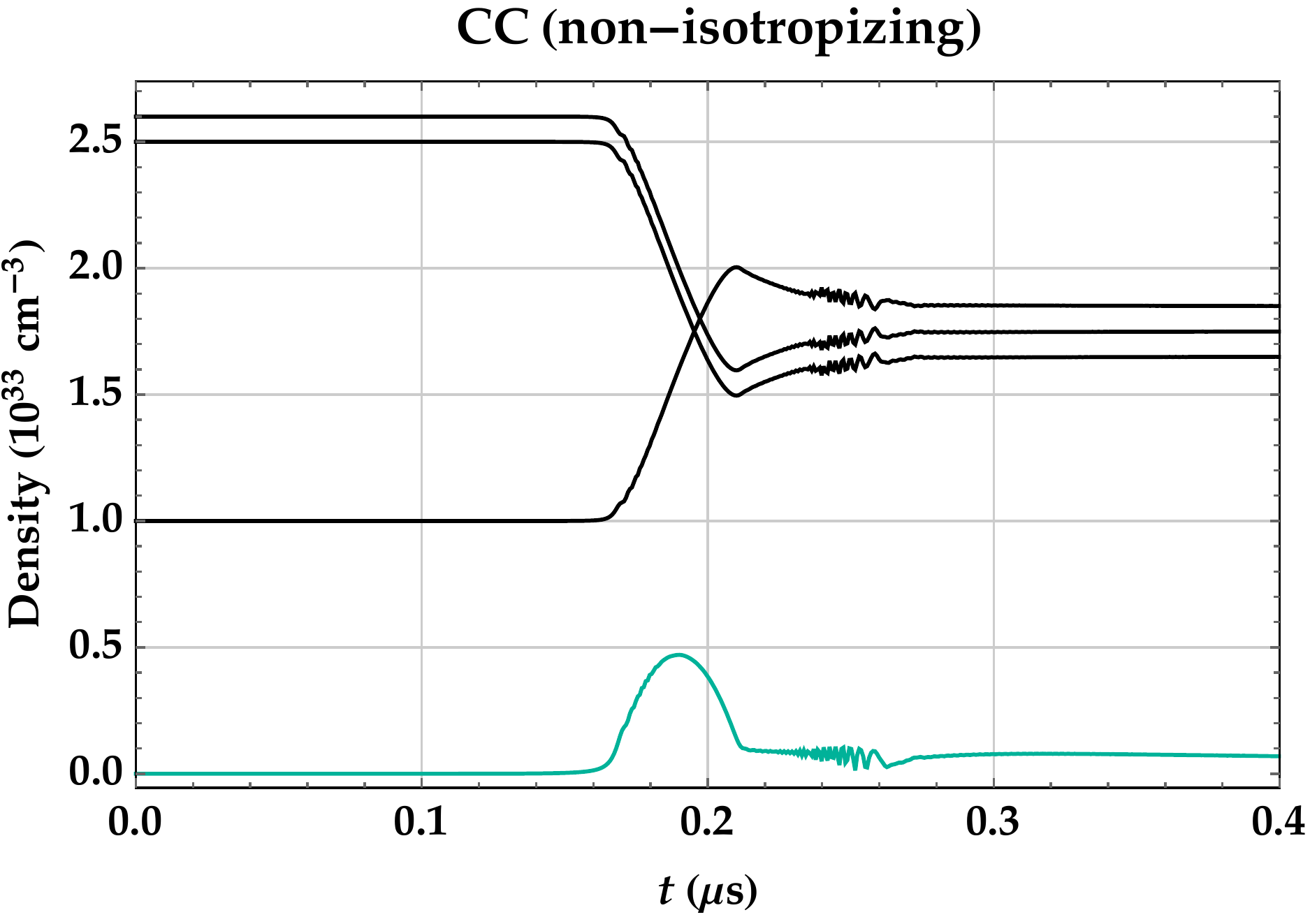}
}
\end{subfigure}

\begin{subfigure}{
\centering
\includegraphics[width=.310\textwidth]{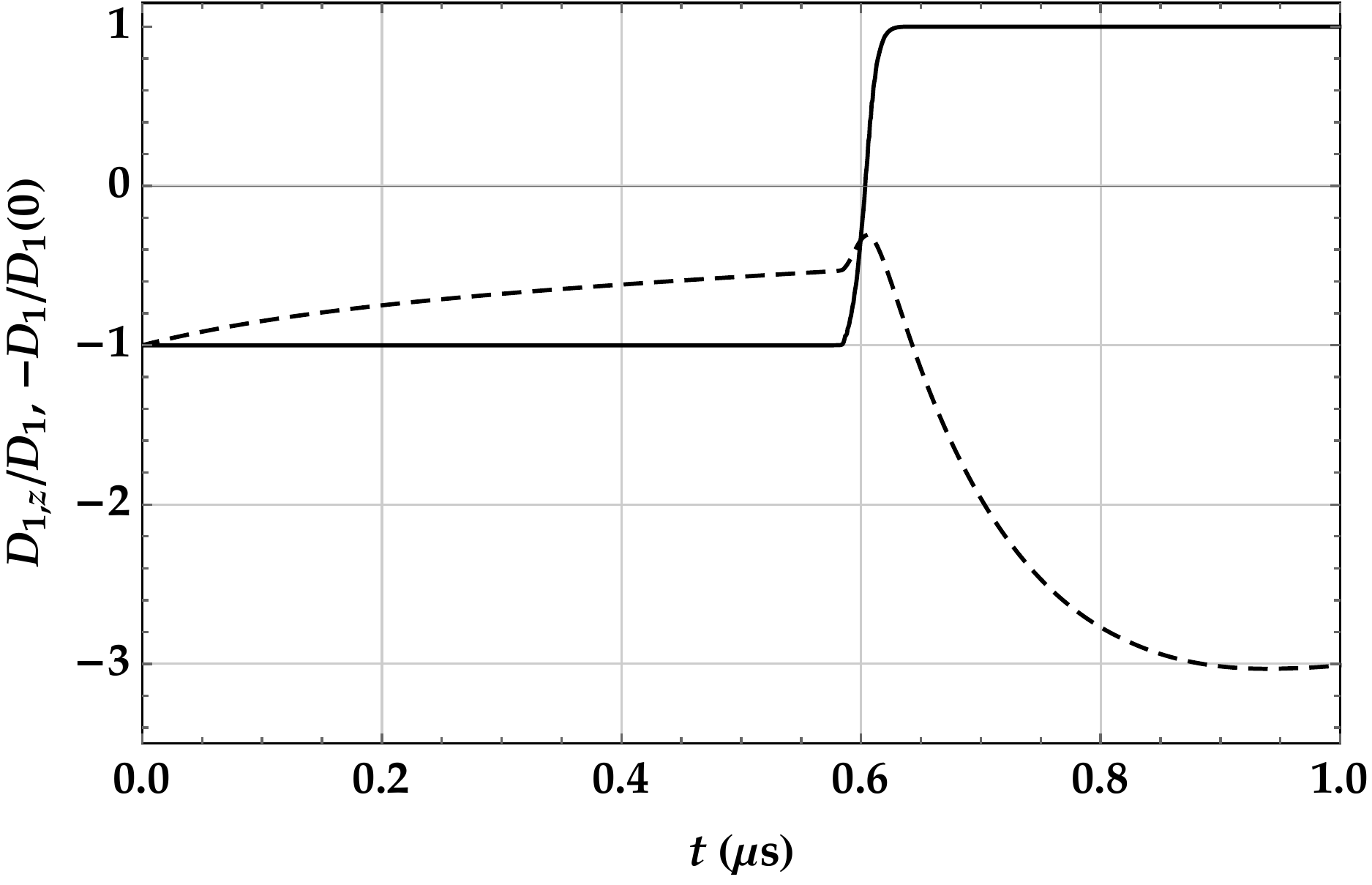}
}
\end{subfigure}
\begin{subfigure}{
\centering
\includegraphics[width=.310\textwidth]{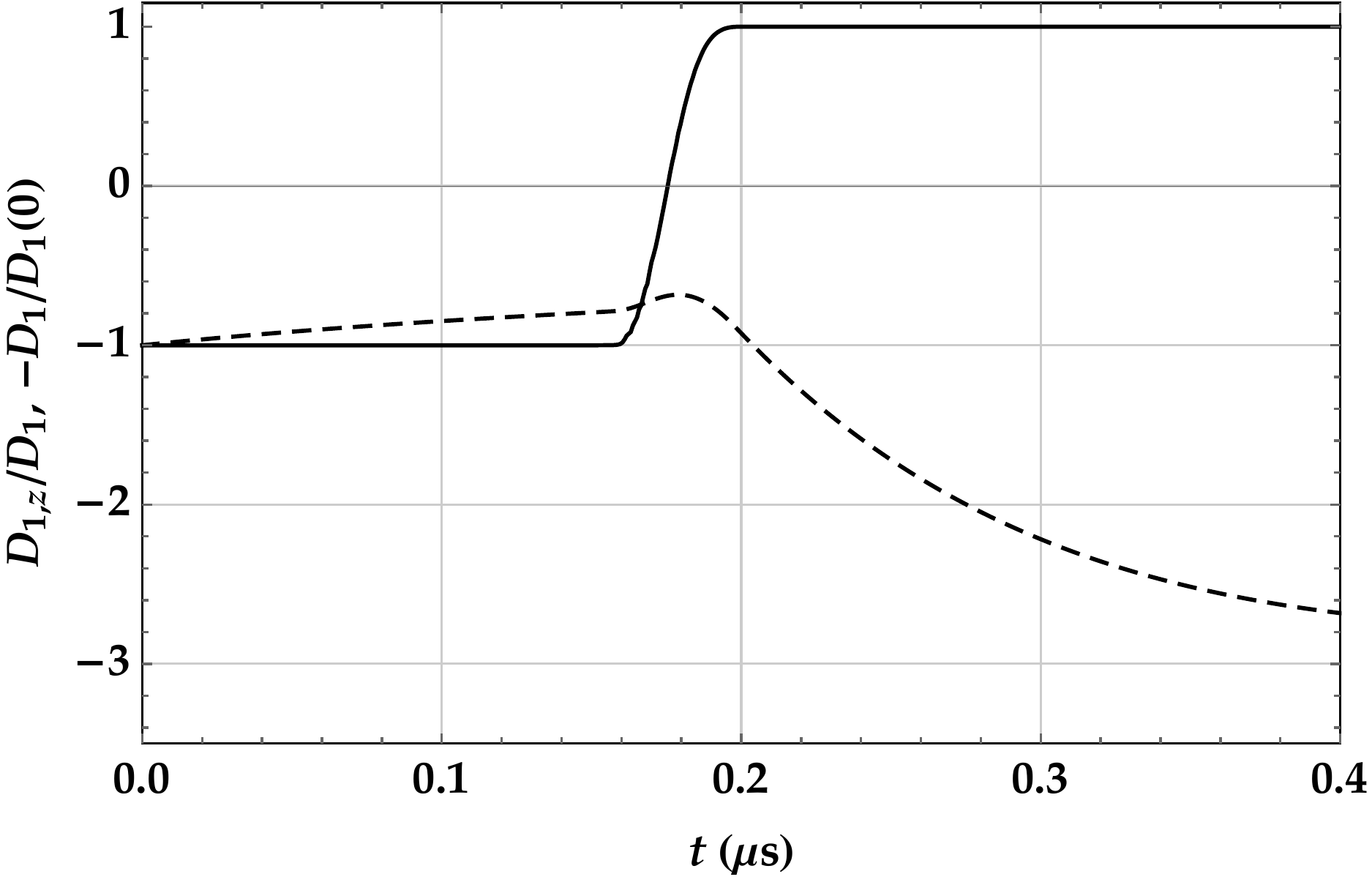}
}
\end{subfigure}
\begin{subfigure}{
\centering
\includegraphics[width=.310\textwidth]{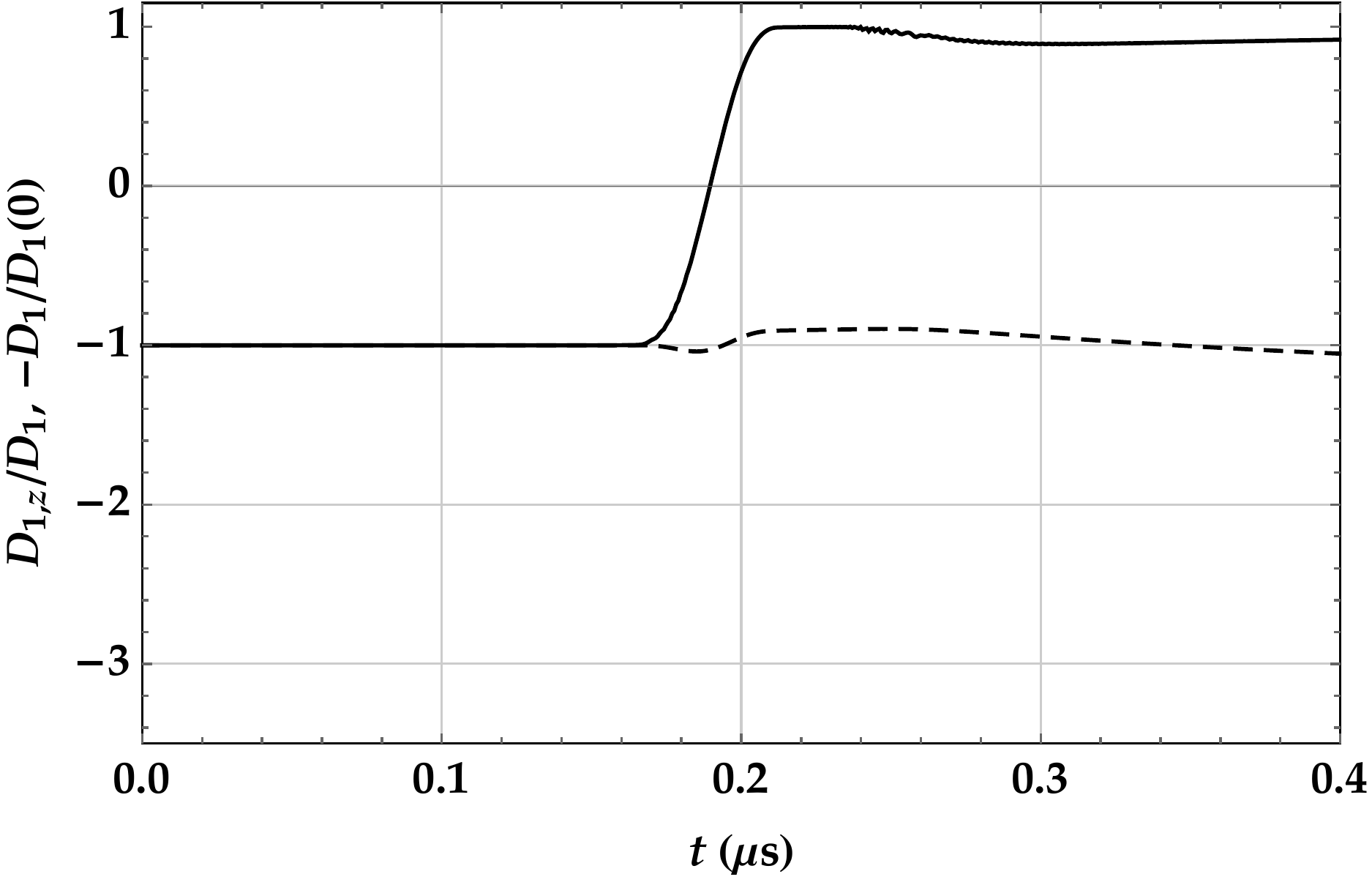}
}
\end{subfigure}

\begin{subfigure}{
\centering
\includegraphics[width=.310\textwidth]{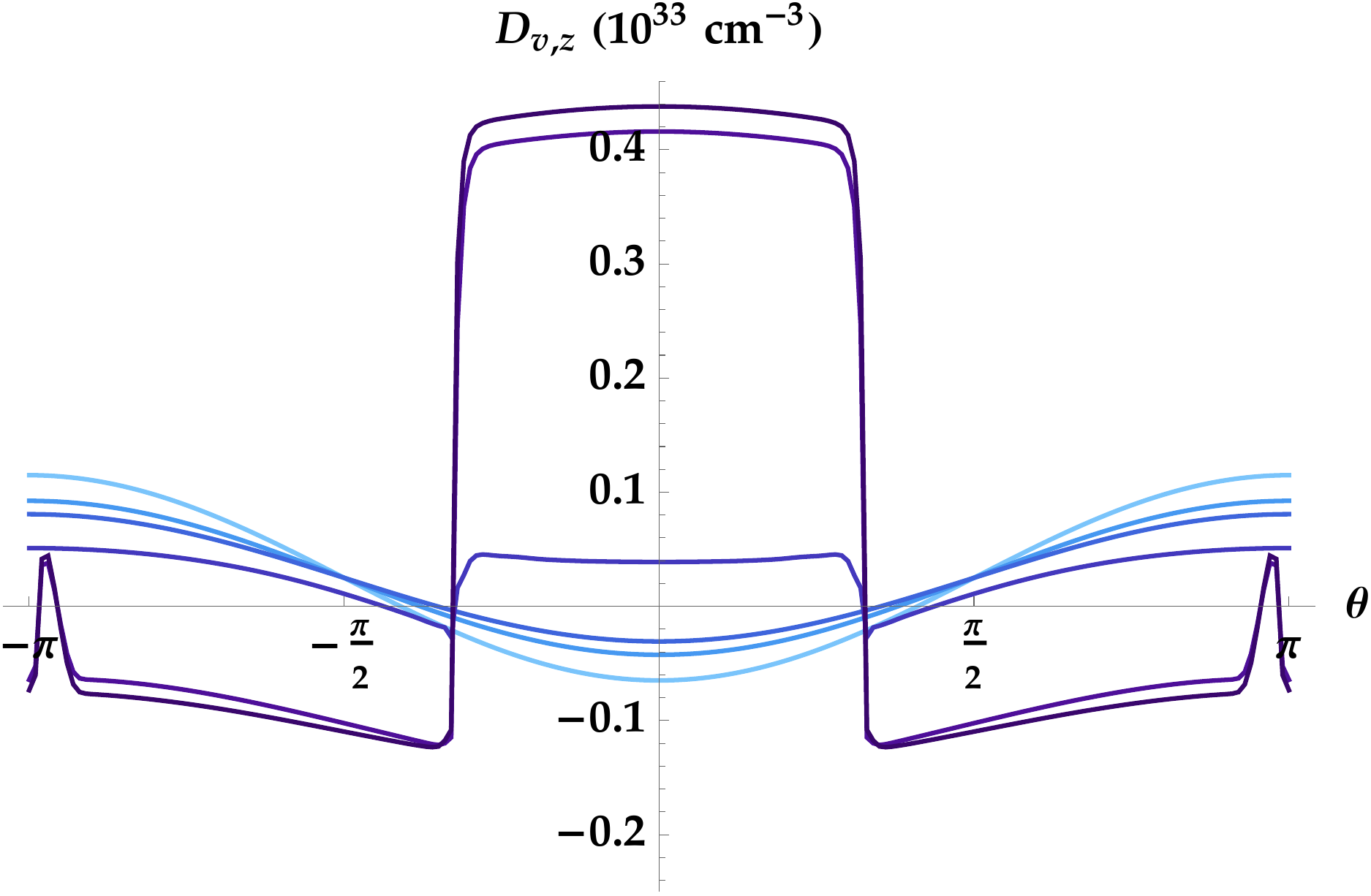}
}
\end{subfigure}
\begin{subfigure}{
\centering
\includegraphics[width=.310\textwidth]{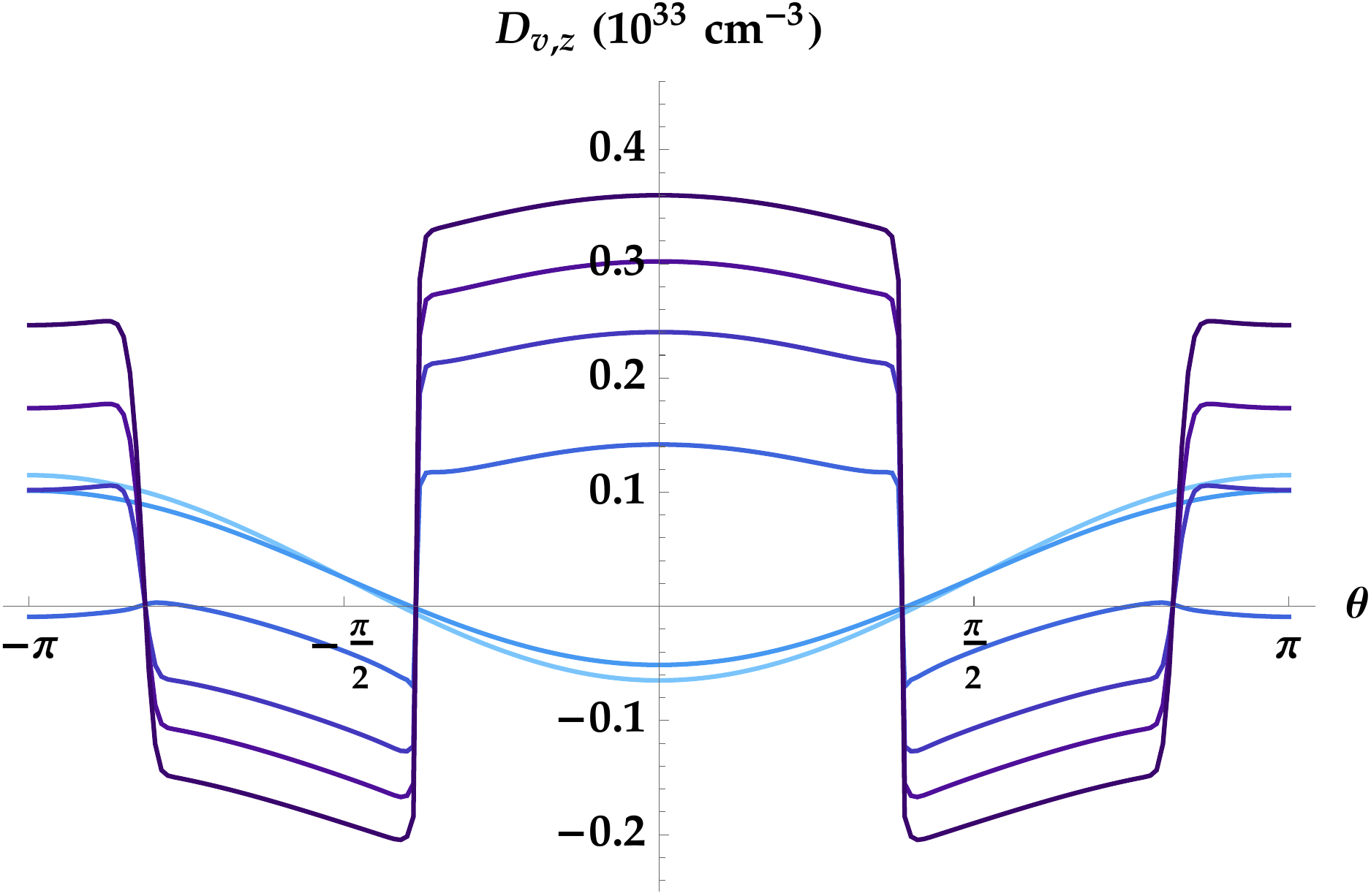}
}
\end{subfigure}
\begin{subfigure}{
\centering
\includegraphics[width=.310\textwidth]{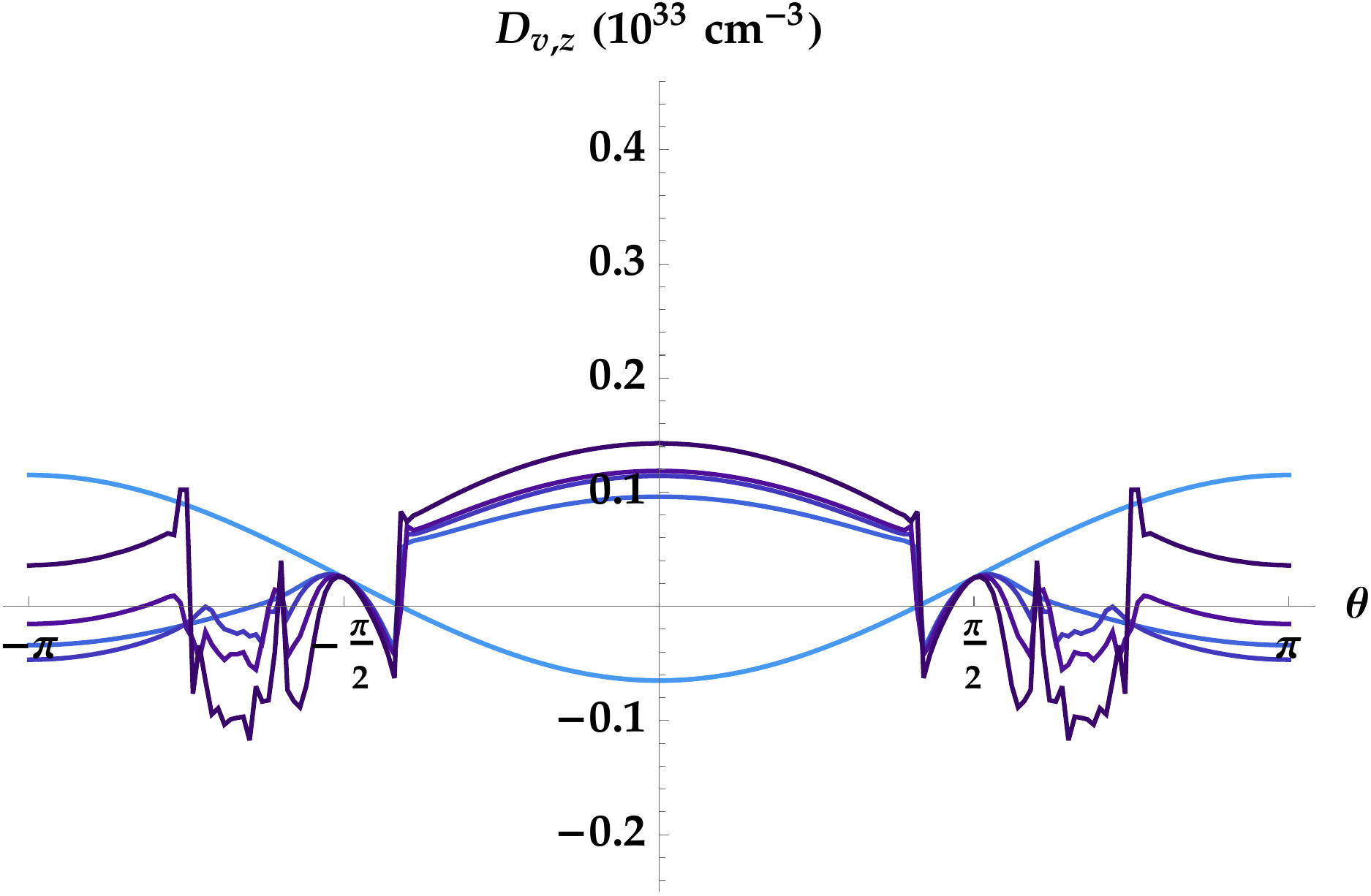}
}
\end{subfigure}

\begin{subfigure}{
\centering
\includegraphics[width=.310\textwidth]{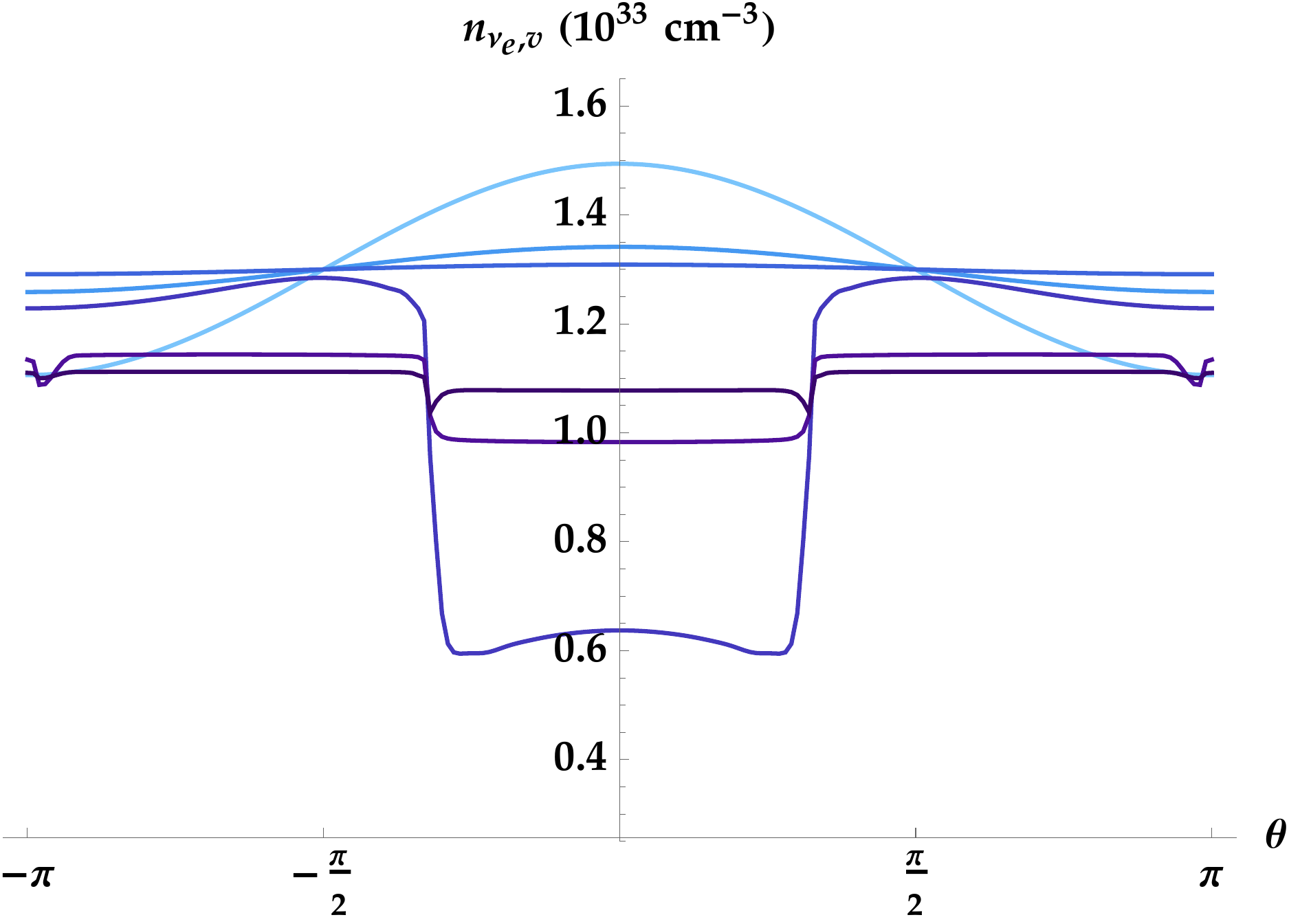}
}
\end{subfigure}
\begin{subfigure}{
\centering
\includegraphics[width=.310\textwidth]{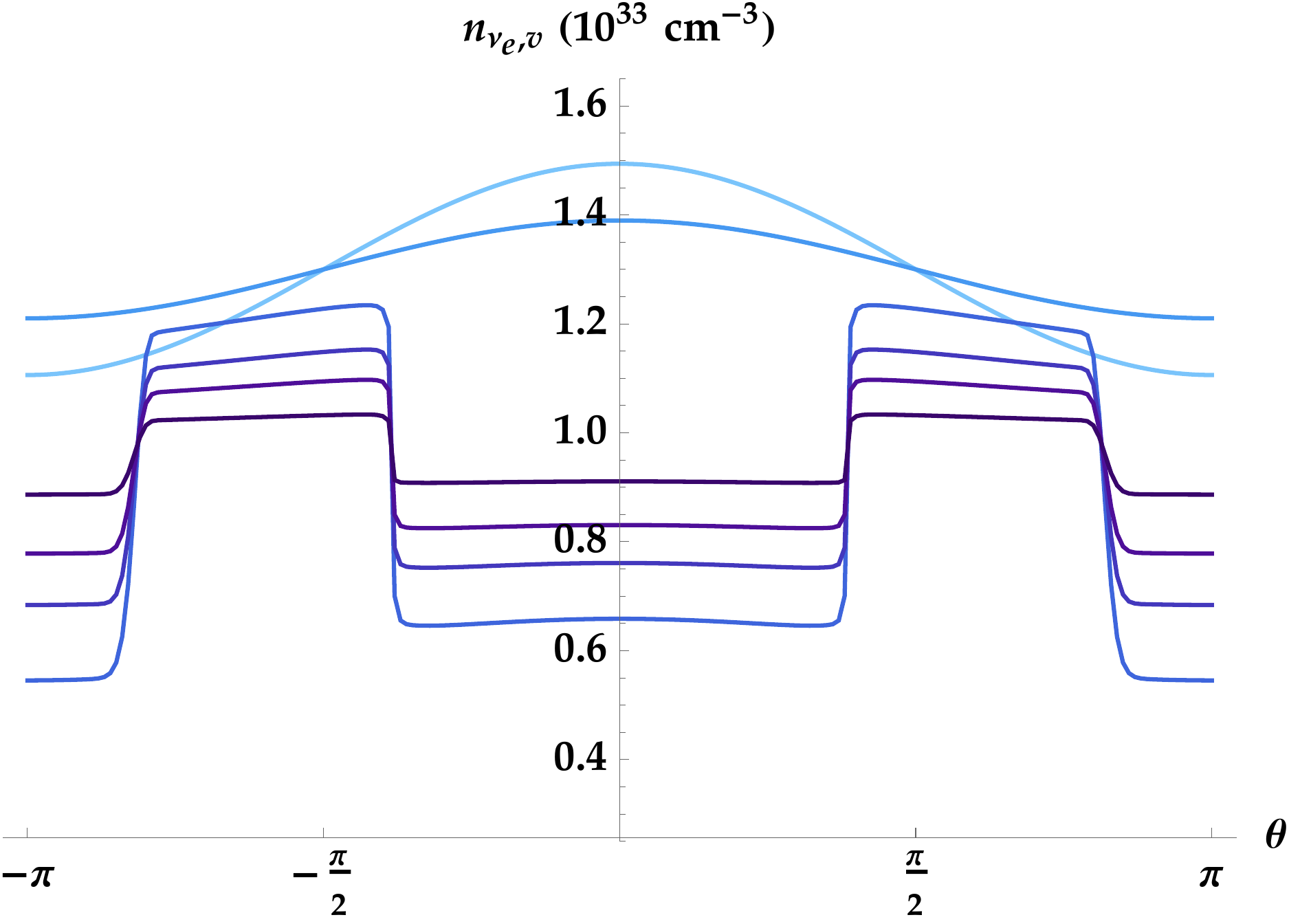}
}
\end{subfigure}
\begin{subfigure}{
\centering
\includegraphics[width=.310\textwidth]{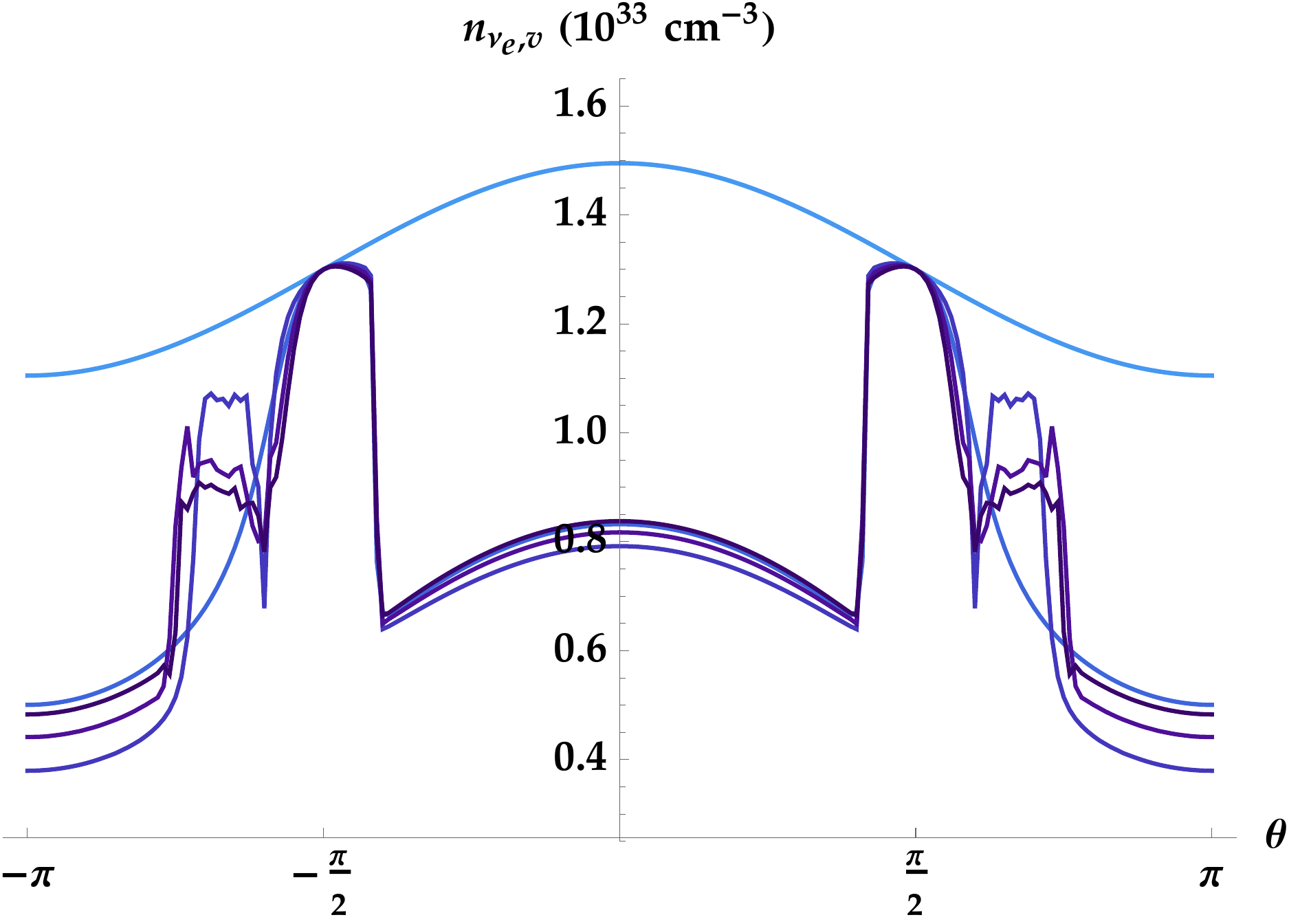}
}
\end{subfigure}
\caption{Collisionally unstable flavor evolution \textit{without} fast instability [angular distributions are given by Eq.~\eqref{eq:ang_noffc}] in the \textit{inverted} mass hierarchy. The quantities plotted are the same as in Fig.~\ref{noffc_nh}. The differences with that figure (which has all the same parameters but assumes the normal hierarchy) are small, indicating that $\omega \neq 0$ effects---and in particular slow instabilities---are not responsible for the main dynamical features witnessed in the two figures. This is as expected for evolution driven by collisional instability.}  
\label{noffc_ih}
\end{figure*}

In this section we employ the angular distributions shown in the top panel of Fig.~\ref{ang_distrs}. The system is stable to fast instability but unstable to collisional instability.

Results are presented in Figs.~\ref{noffc_nh} and \ref{noffc_ih}. The calculations underlying these figures use identical parameters except for the choice of the neutrino mass hierarchy: the first is normal (NH) and the second is inverted (IH). The differences between Figs.~\ref{noffc_nh} and \ref{noffc_ih} are minor. The comparison is useful nonetheless because it rules out the possibility that, for example, the multi-zenith-angle (MZA) slow instability \cite{raffelt2007b} is responsible for the observed dynamics. We discuss this point in more detail below.

While there is variation in the timing and details, all three implementations of scattering (NC, isotropizing CC, and non-isotropizing CC) exhibit collisional instability, as seen in the top rows of Figs.~\ref{noffc_nh} and \ref{noffc_ih}. In calculations with $\Gamma = \bar{\Gamma}$ (not displayed here), all plotted quantities remain nearly constant. These observations alone extends the numerical findings of Ref.~\cite{johns2021b}, where the dominant processes were CC absorption and emission.

The fact that collisional instability can be caused by NC scattering is especially interesting because this process does not cause decoherence at the single-particle level. Indeed, in the isotropic model of Ref.~\cite{johns2021b}, NC scattering has no effect at all on the dynamics. Yet decoherence is the key to collisional instability. We see here that the decoherence that comes from coherently averaging over flavor states (\textit{i.e.}, in the gain term $\langle \rho \rangle$) is capable of collisionally destabilizing the system. From a certain viewpoint, CC interactions induce decoherence through system--environment coupling and NC scattering through system--system coupling. It turns out that collisional instability can result from either one.

While the overall finding of collisional instability is consistent across all three scattering implementations, the quantitative details differ from one implementation to the next. The instability develops most slowly under the influence of NC scattering, and the overall flavor conversion is the least in this case. With non-isotropizing CC scattering, some residual flavor coherence is left at $t = 0.4~\mu s$, the final time plotted, and in this case flavor conversion is at its greatest.

The second rows of Figs.~\ref{noffc_nh} and \ref{noffc_ih} plot the evolution of the difference vector $\mathbf{D}_1$, which is formed in the following way. Define the polarization vectors $\mathbf{P}_v$ and $\mathbf{\bar{P}}_v$ by
\begin{align}
&\rho_v = P_{v, 0} + \mathbf{P}_v \notag \\
&\bar{\rho}_v = \bar{P}_{v, 0} + \mathbf{\bar{P}}_v,
\end{align}
such that the traces $2P_{v,0}$ and $2\bar{P}_{v,0}$ are the number densities summed over flavor. Let $l$ denote the $l$th Legendre moment, as in
\begin{equation}
\rho_{l} = \int_{-1}^{+1} dv L_l(v) \rho_v.
\end{equation}
Then $\mathbf{D}_1$ is the $l = 1$ moment of $\mathbf{D}_v = \mathbf{P}_v - \mathbf{\bar{P}}_v$:
\begin{equation}
\mathbf{D}_1 = \int_{-1}^{+1} dv ~v ( \mathbf{P}_v - \mathbf{\bar{P}}_v ).
\end{equation}
Also important is the monopole difference vector
\begin{equation}
\mathbf{D}_0 = \int_{-1}^{+1} dv ( \mathbf{P}_v - \mathbf{\bar{P}}_v ), \label{eq:D0}
\end{equation}
whose initial magnitude was used in Table~\ref{paramtab} to define the strength of the self-coupling potential.

For all three implementations of collisions, the instability is characterized by $\mathbf{D}_1$ inverting its position. By contrast, $\mathbf{D}_0$ (not plotted) is always nearly constant in these calculations. The collisional instabilities in this paper are distinct from those of Ref.~\cite{johns2021b} in that they are intrinsically anisotropic, here being facilitated by the dipole part of $H^{\nu\nu}$.

The vector $\mathbf{D}_1$ is also critical to fast instabilities \cite{johns2020, padillagay2021} and MZA instabilities \cite{raffelt2007b, johns2020b}. Since the system begins near the threshold of fast instability, a concern in interpreting our results is that fast instability might be generated dynamically. This appears not to be the case. In the isotropizing calculations, the magnitude $D_1$ shrinks in the lead-up to the instability, which moves the system more deeply into the region of parameter space that is stable to fast oscillations. In the non-isotropizing calculations, there is minuscule change in $D_{v,z}$ until the instability sets in. Moreover, we have checked that a similar inversion of $D_{1,z}$ occurs in calculations with only three moments, for which FFC is not possible. A related concern is that the MZA instability is coming into play. The comparison of mass hierarchies confirms that this instability is not relevant here either. Thus the calculations in Figs.~\ref{noffc_nh} and \ref{noffc_ih} are isolating the collisional instabilities, as desired.

We can analytically observe how anisotropic collisional instabilities appear by noting that $\mathbf{D}_1$ has equation of motion
\begin{equation}
\dot{\mathbf{D}}_1 = \omega \mathbf{B} \times \mathbf{S}_1 + \frac{2}{3} \mu \left( 2 \mathbf{D}_0 + \mathbf{D}_2 \right) \times \mathbf{D}_1 - \Gamma_+ \mathbf{D}_{1,T} - \Gamma_- \mathbf{S}_{1,T},
\end{equation}
where
\begin{equation}
\Gamma_{\pm} = \frac{\Gamma \pm \bar{\Gamma}}{2}.
\end{equation}
Taking $\omega \ll \mu, \Gamma, \bar{\Gamma}$, we have
\begin{equation}
\dot{\mathbf{D}}_1 = \mathbf{H}_1 \times \mathbf{D}_1 - \Gamma_+ \mathbf{D}_{1,T} - \Gamma_- \mathbf{S}_{1,T}.
\end{equation}
In introducing $\mathbf{H}_1 \equiv (2\mu / 3) ( 2 \mathbf{D}_0 + \mathbf{D}_2 )$, we are emphasizing that the collisionless part of the evolution is just precession around a (time-dependent) Hamiltonian vector. If $\mathbf{D}_1$, $\mathbf{S}_1$, and $\mathbf{H}_1$ are all nearly aligned or antialigned with $\mathbf{z}$, as they are initially, then the cross product is small and we can write
\begin{equation}
\dot{\mathbf{D}}_{1,T} \cong \left( - \Gamma_+ \mp \frac{| \mathbf{S}_1 |}{| \mathbf{D}_1 |} \Gamma_- \right) \mathbf{D}_{1,T},
\end{equation}
where the choice of sign must be consistent with $\mathbf{\hat{S}}_1 \sim \pm \mathbf{\hat{D}}_1$. An exponentially growing solution is possible if $\mathbf{D}_1$ and $\mathbf{S}_1$ are nearly antialigned.

This instability is very similar to a collisional instability in $\mathbf{D}_0$ that was pointed out in Ref.~\cite{johns2021b}. While the $\mathbf{D}_0$ instability is probably not relevant to supernovae (note that Ref.~\cite{johns2021b} identified an additional monopole collisional instability that \textit{is} potentially relevant), the $\mathbf{D}_1$ instability is plausible because we typically have a hierarchy of number densities
\begin{equation}
n_{\nu_e} > n_{\bar{\nu}_e} > n_{\nu_x}
\end{equation}
and a hierarchy of flux factors
\begin{equation}
f_{\nu_x} > f_{\bar{\nu}_e} > f_{\nu_e}.
\end{equation}
Indeed, with the parameters we have chosen, $\mathbf{z} \cdot \mathbf{\hat{S}}_1 (0) = +1$ and $\mathbf{z} \cdot \mathbf{\hat{D}}_1 (0) = -1$. Accordingly, collisions cause $\mathbf{D}_1$ to flips its orientation in all of the results we present.

The bottom two rows of Figs.~\ref{noffc_nh} and \ref{noffc_ih} show $D_{v,z} (t)$ and $n_{\nu_e, v} (t)$. The $\mathbf{D}_1$ inversion is reflected here. In the left two columns, features flatten out over time as the angular distributions isotropize.

Once the instability sets in, the curves all change sharply at the location of the initial angular crossing, \textit{i.e.}, the direction $v$ at which $D_{v,z}(0) = 0$. For non-isotropizing CC scattering, narrow dips freeze in just behind this location. In all cases, when the collisional instability is first developing, it manifests as dips right at the angular crossings. This can be understood using the resonant-trajectory analysis of Ref.~\cite{johns2020} (see, in that paper, Eq. 10 and the dips in the top row of Fig. 2). According to that analysis, with the $l \geq 2$ moments vanishing, the resonant trajectory and the crossing location coincide. In the isotropizing calculations, the dips broaden as time progresses.

Because the neutrino--neutrino Hamiltonian has the form $\mathbf{H}_v^{\nu\nu} \propto \mathbf{D}_0 - v \mathbf{D}_1$, the dynamics of $\mathbf{D}_1$ does not influence the flavor states at $v = 0$ ($\theta = \pi / 2$). In the plots with non-isotropizing CC scattering, these points are frozen at their initial values.

Still focusing on the non-isotropizing case, we observe ripples appearing beyond $\theta = \pm \pi / 2$ at later times. Additional crossings in $D_{v,z}$ are generated as $\mathbf{D}_1$ flips. Ripples start at these crossings and spread outward. We hypothesize that the reason may be the following. For a given trajectory $v$, we have
\begin{align}
&\dot{\mathbf{S}}_v = \omega \mathbf{B} \times \mathbf{D}_v + \mu \left( \mathbf{D}_0 - v \mathbf{D}_1 \right) \times \mathbf{S}_v \notag \\
& \hspace{0.9 in} - \Gamma_+ \mathbf{S}_{v,T} - \Gamma_- \mathbf{D}_{v,T} \notag \\
&\dot{\mathbf{D}}_v = \omega \mathbf{B} \times \mathbf{S}_v + \mu \left( \mathbf{D}_0 - v \mathbf{D}_1 \right) \times \mathbf{D}_v \notag \\
& \hspace{0.9 in} - \Gamma_+ \mathbf{D}_{v,T} - \Gamma_- \mathbf{S}_{v,T}.
\end{align}
As before, we consider the limit of small $\omega$. We assume further that the relevant vectors are all close enough to being (anti)aligned that the cross products with $H^{\nu\nu}_v$ can be ignored. Then, dropping the oscillation terms,
\begin{align}
&\dot{\mathbf{S}}_{v,T} = - \Gamma_+ \mathbf{S}_{v,T} - \Gamma_- \mathbf{D}_{v,T}, \notag \\
&\dot{\mathbf{D}}_{v,T} = - \Gamma_+ \mathbf{D}_{v,T} - \Gamma_- \mathbf{S}_{v,T}.
\end{align}
If
\begin{equation}
\frac{\Gamma_-}{\Gamma_+} \gtrsim \frac{ | \mathbf{D}_v |}{| \mathbf{S}_v |},
\end{equation}
then $\mathbf{S}_v$ is collisionally unstable due to the growing solution of
\begin{equation}
\ddot{\mathbf{S}}_{v,T} + \Gamma_+ \dot{\mathbf{S}}_{v,T} - \Gamma_-^2 \mathbf{S}_{v,T} = 0.
\end{equation}
This analysis is once again very similar to the one given in Ref.~\cite{johns2021b} for the isotropic collisional instability, but here a single propagation direction is being picked out. The instability is most likely to occur where $\mathbf{D}_v \sim 0$, \textit{i.e.}, at or near the angular crossings. In this manner the $\mathbf{D}_1$ instability can perhaps generate new angular crossings, which generate new collisional instabilities, which in turn generate ripples. Ripples do \textit{not} form in the dips themselves, however, presumably because these trajectories have already expressed collisional instability.

We offer the analysis in the previous paragraph only as a \textit{possible} mechanism behind the fluctuations, leaving the development of a thorough and verified explanation---of the ripples and of other nuanced features of the numerical results---to future work. The main conclusion we draw in this section is that collisional instability appears in all cases studied.

\section{Collisional instability \textit{with} FFC \label{sec:results2}}

\begin{figure*}
\centering
\begin{subfigure}{
\centering
\includegraphics[width=.310\textwidth]{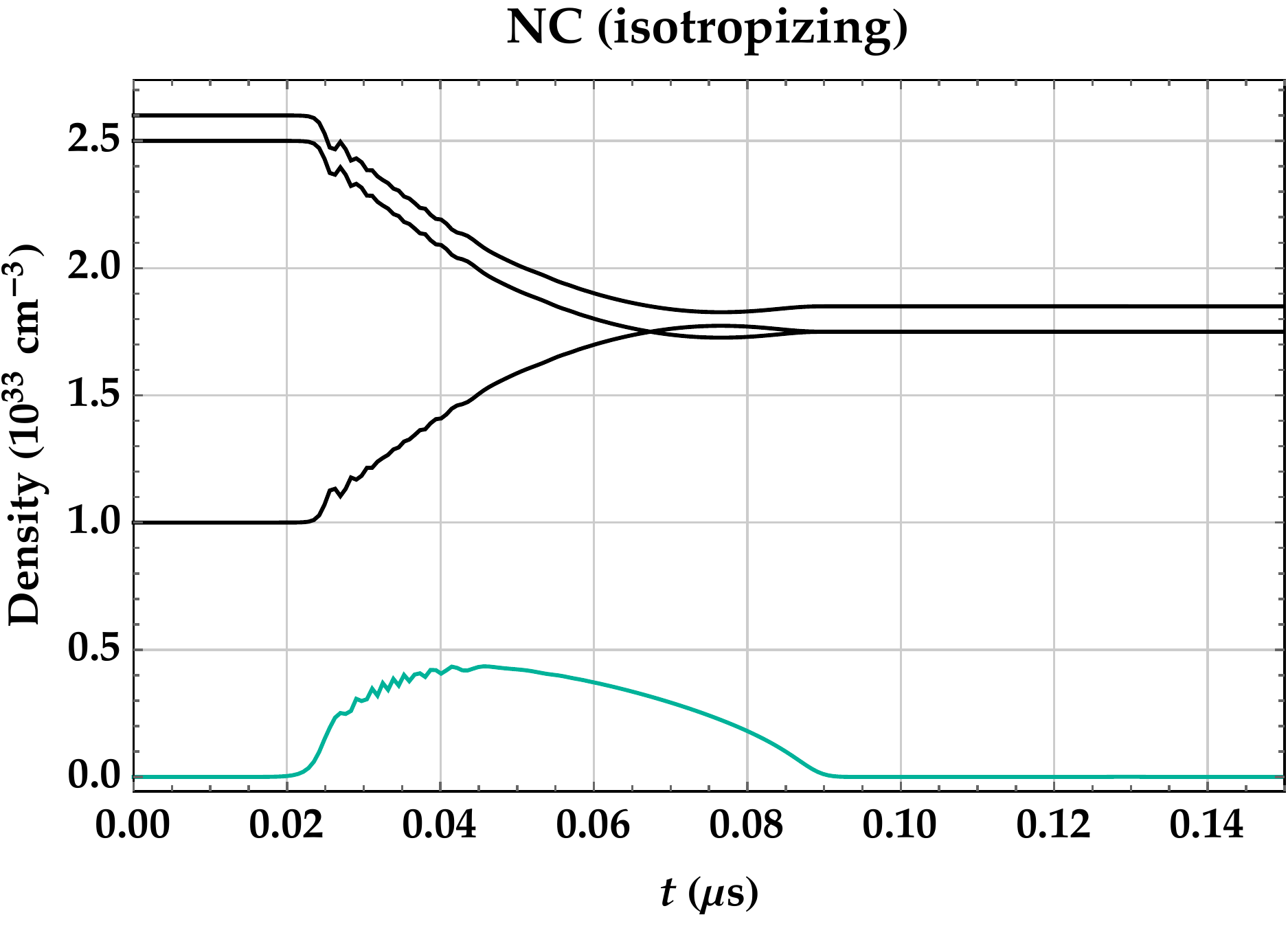}
}
\end{subfigure}
\begin{subfigure}{
\centering
\includegraphics[width=.310\textwidth]{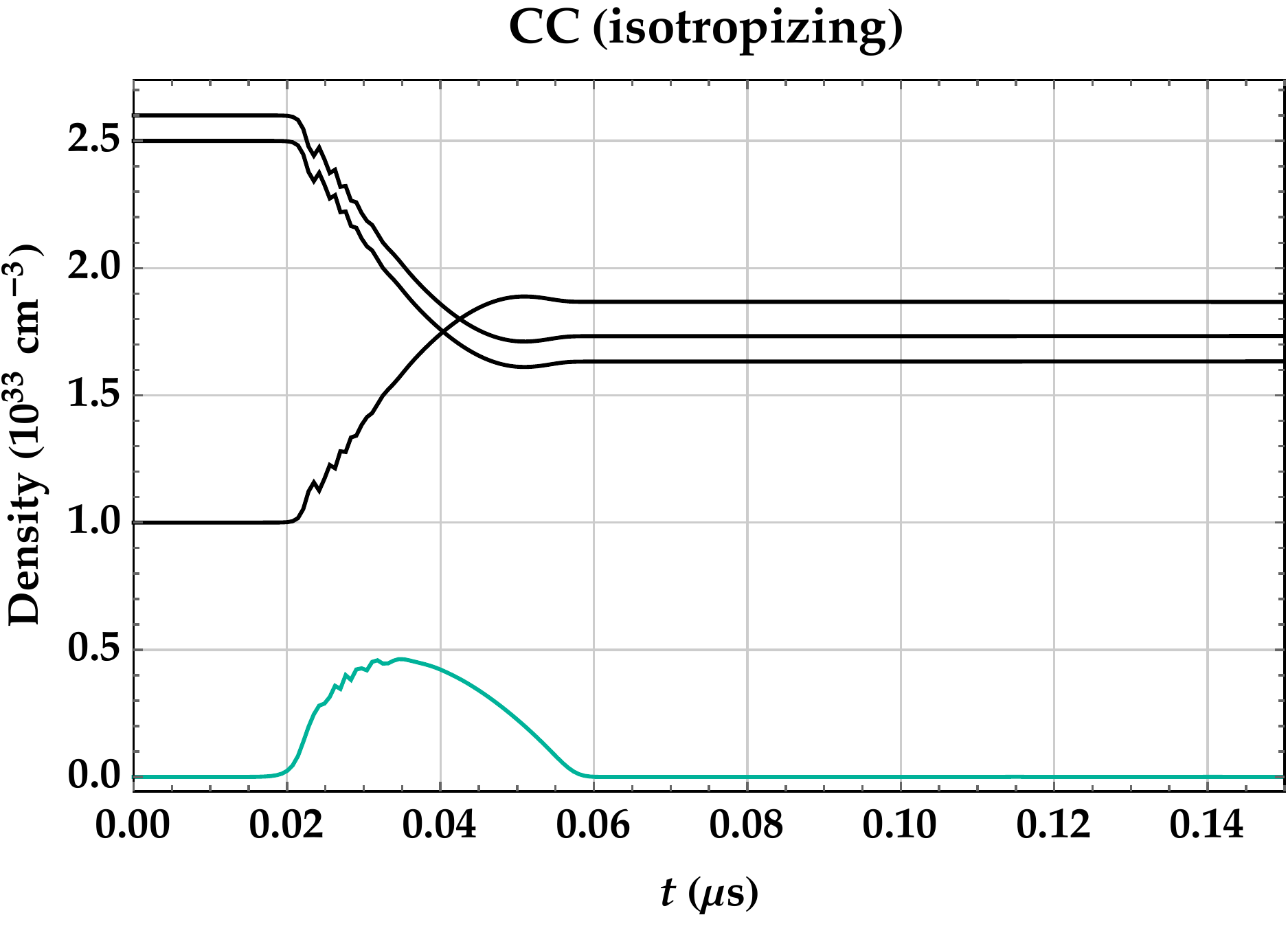}
}
\end{subfigure}
\begin{subfigure}{
\centering
\includegraphics[width=.310\textwidth]{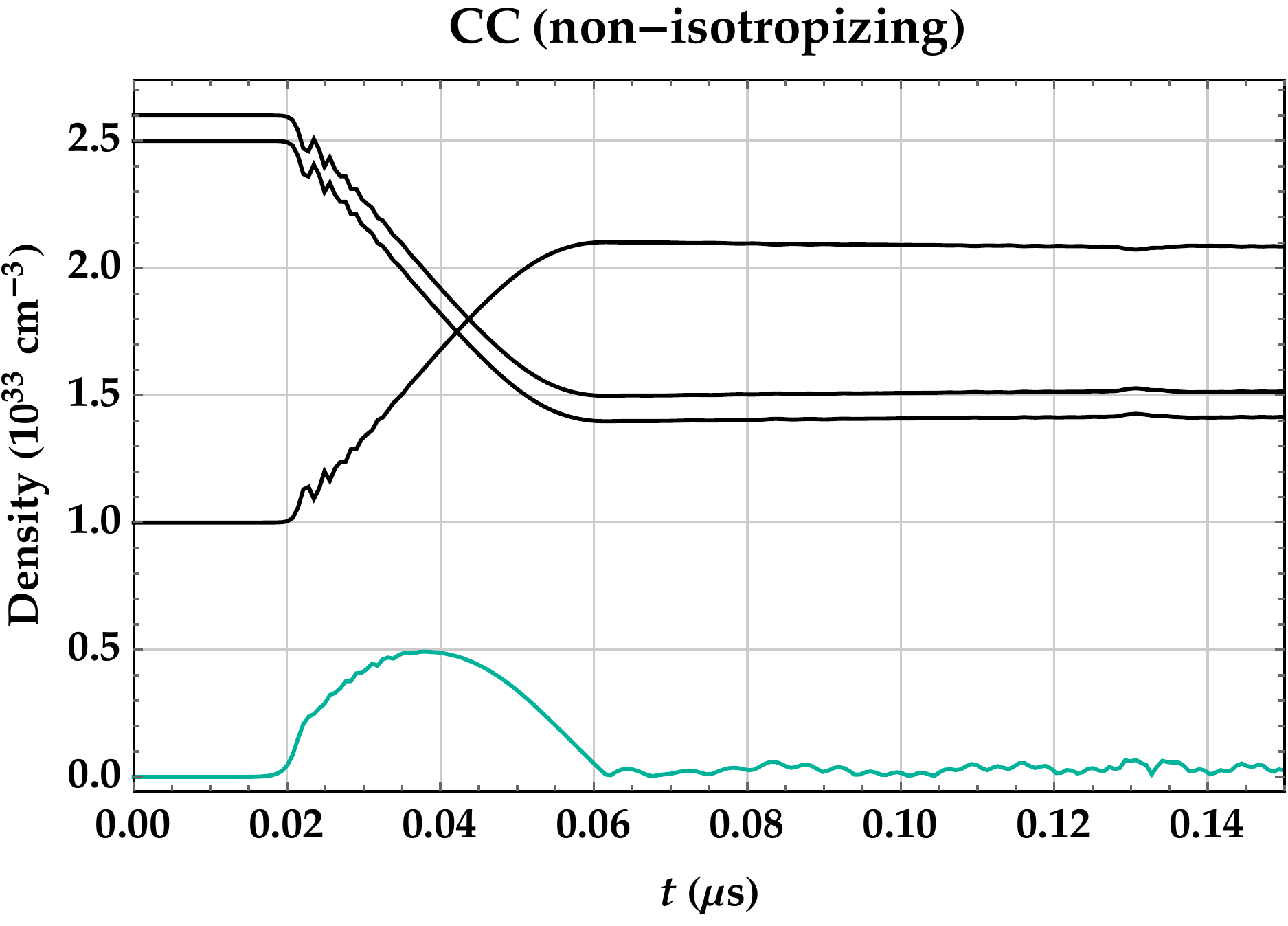}
}
\end{subfigure}

\begin{subfigure}{
\centering
\includegraphics[width=.310\textwidth]{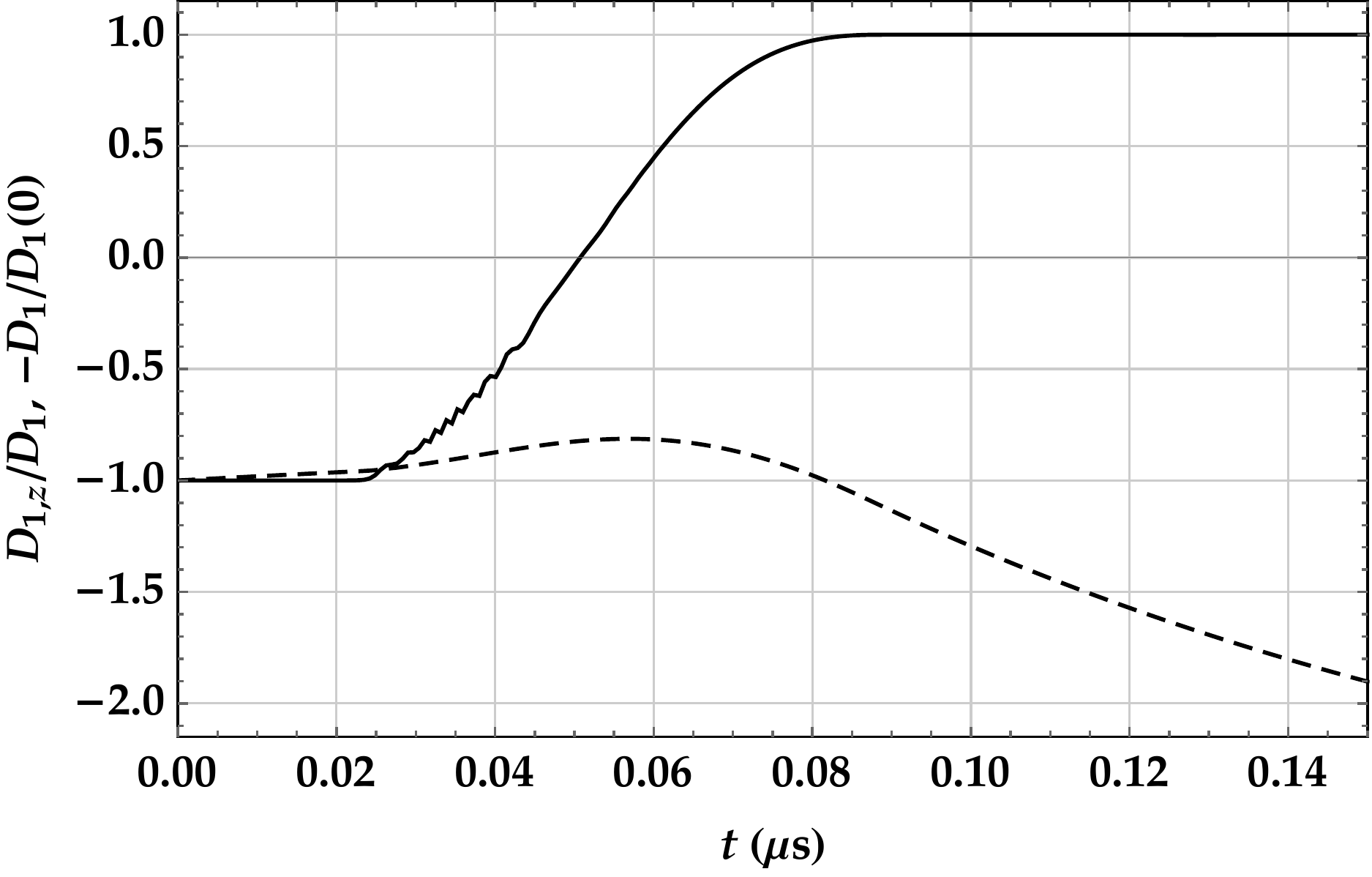}
}
\end{subfigure}
\begin{subfigure}{
\centering
\includegraphics[width=.310\textwidth]{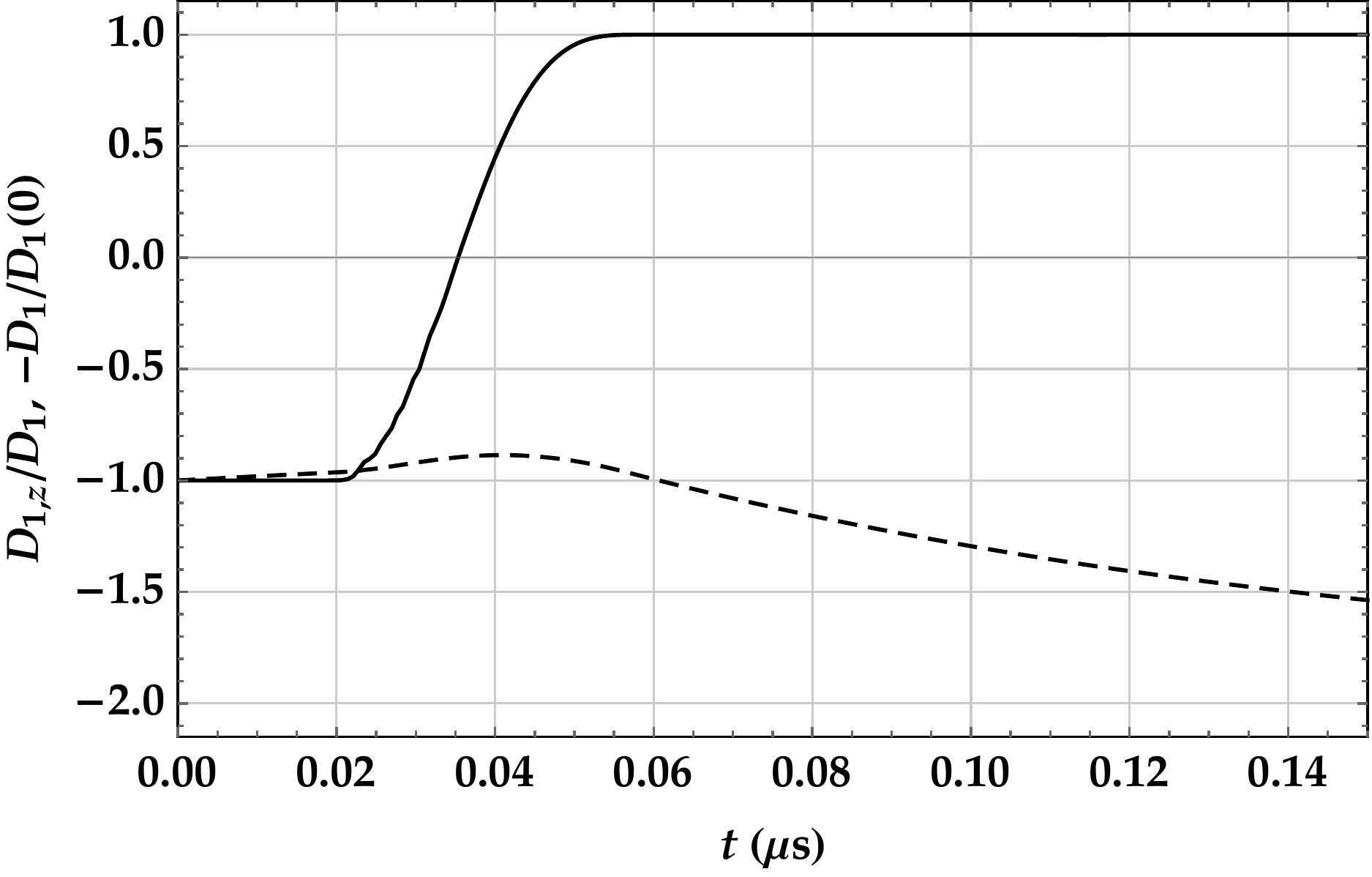}
}
\end{subfigure}
\begin{subfigure}{
\centering
\includegraphics[width=.310\textwidth]{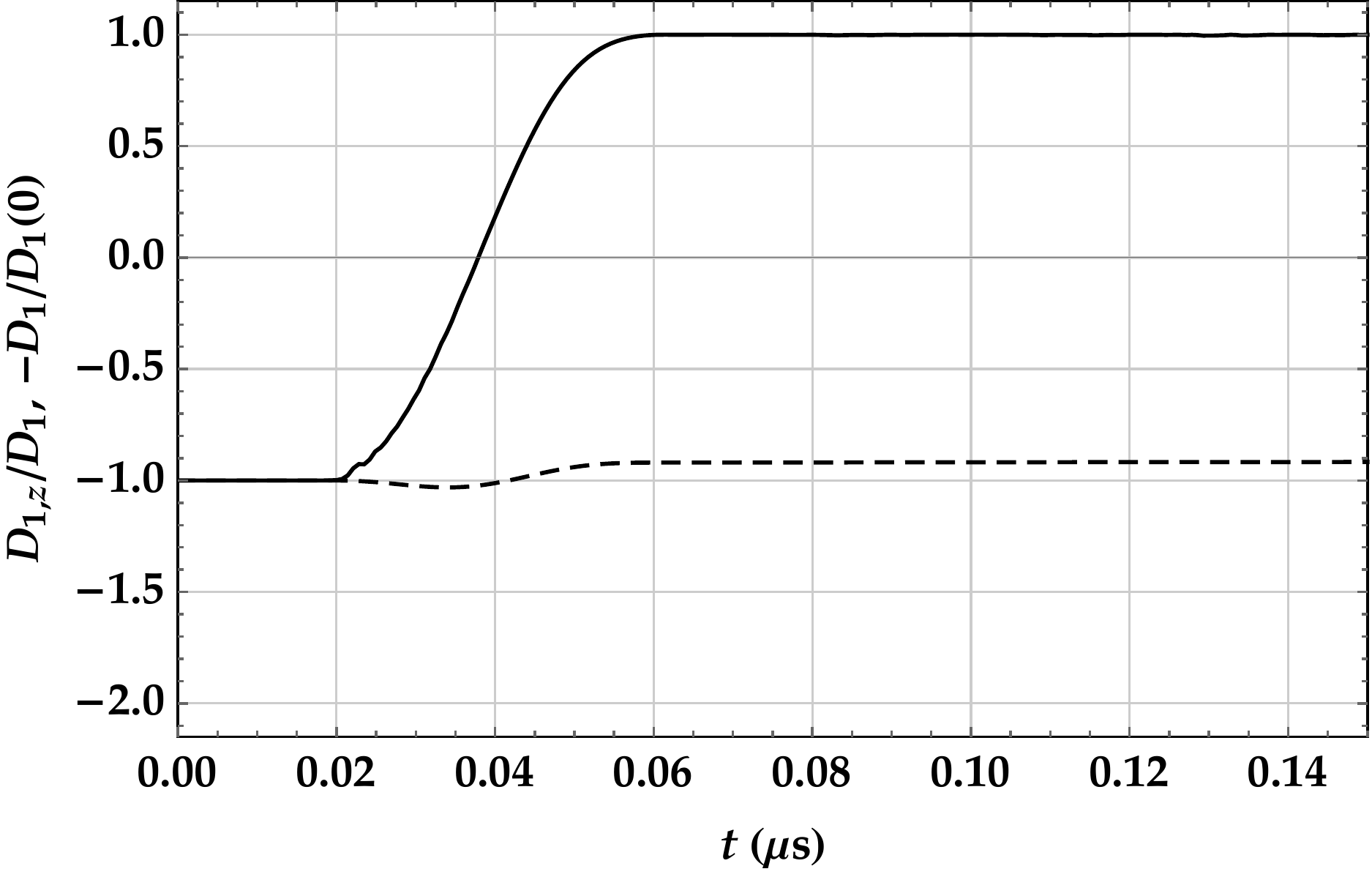}
}
\end{subfigure}

\begin{subfigure}{
\centering
\includegraphics[width=.310\textwidth]{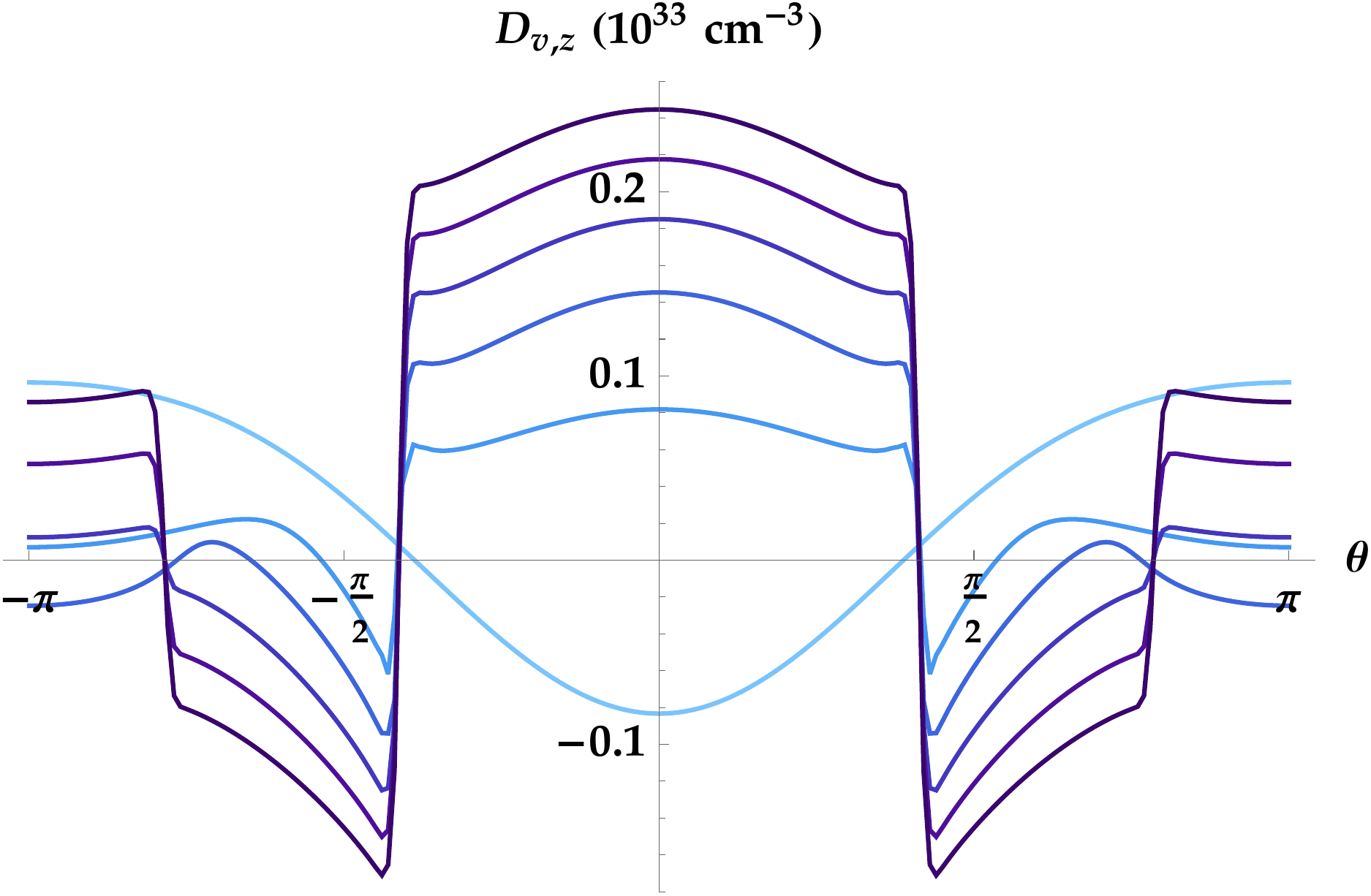}
}
\end{subfigure}
\begin{subfigure}{
\centering
\includegraphics[width=.310\textwidth]{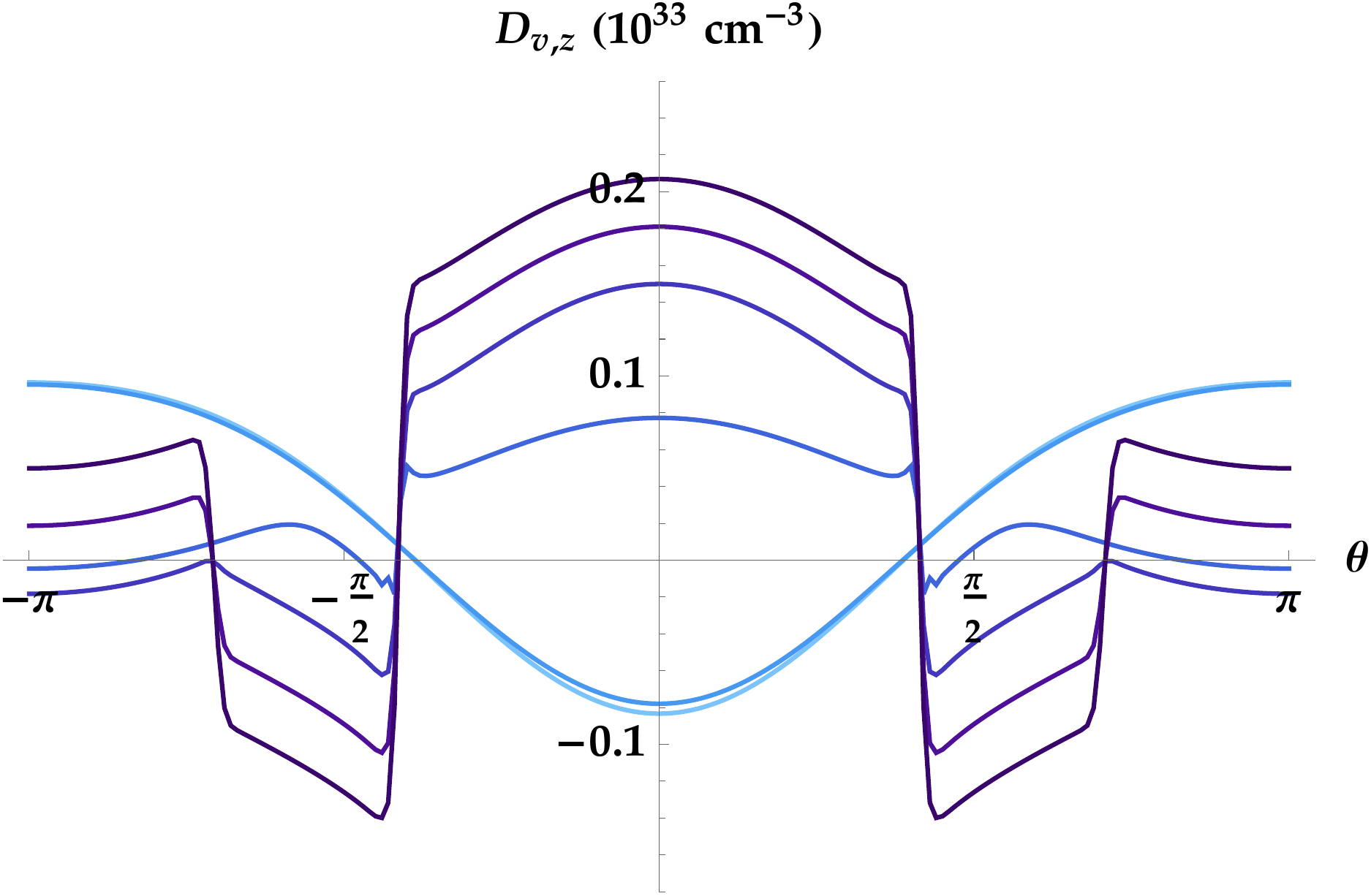}
}
\end{subfigure}
\begin{subfigure}{
\centering
\includegraphics[width=.310\textwidth]{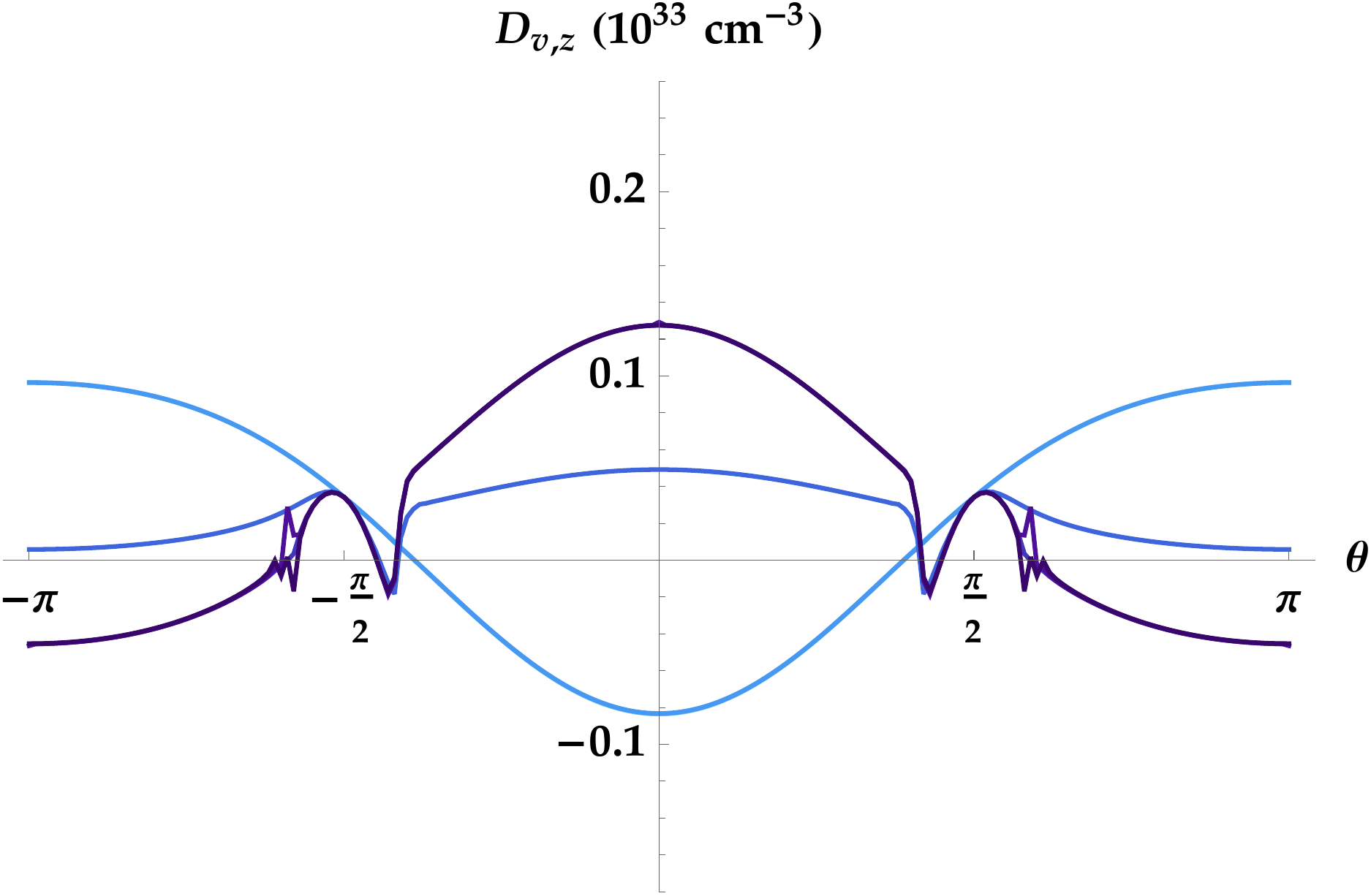}
}
\end{subfigure}

\begin{subfigure}{
\centering
\includegraphics[width=.310\textwidth]{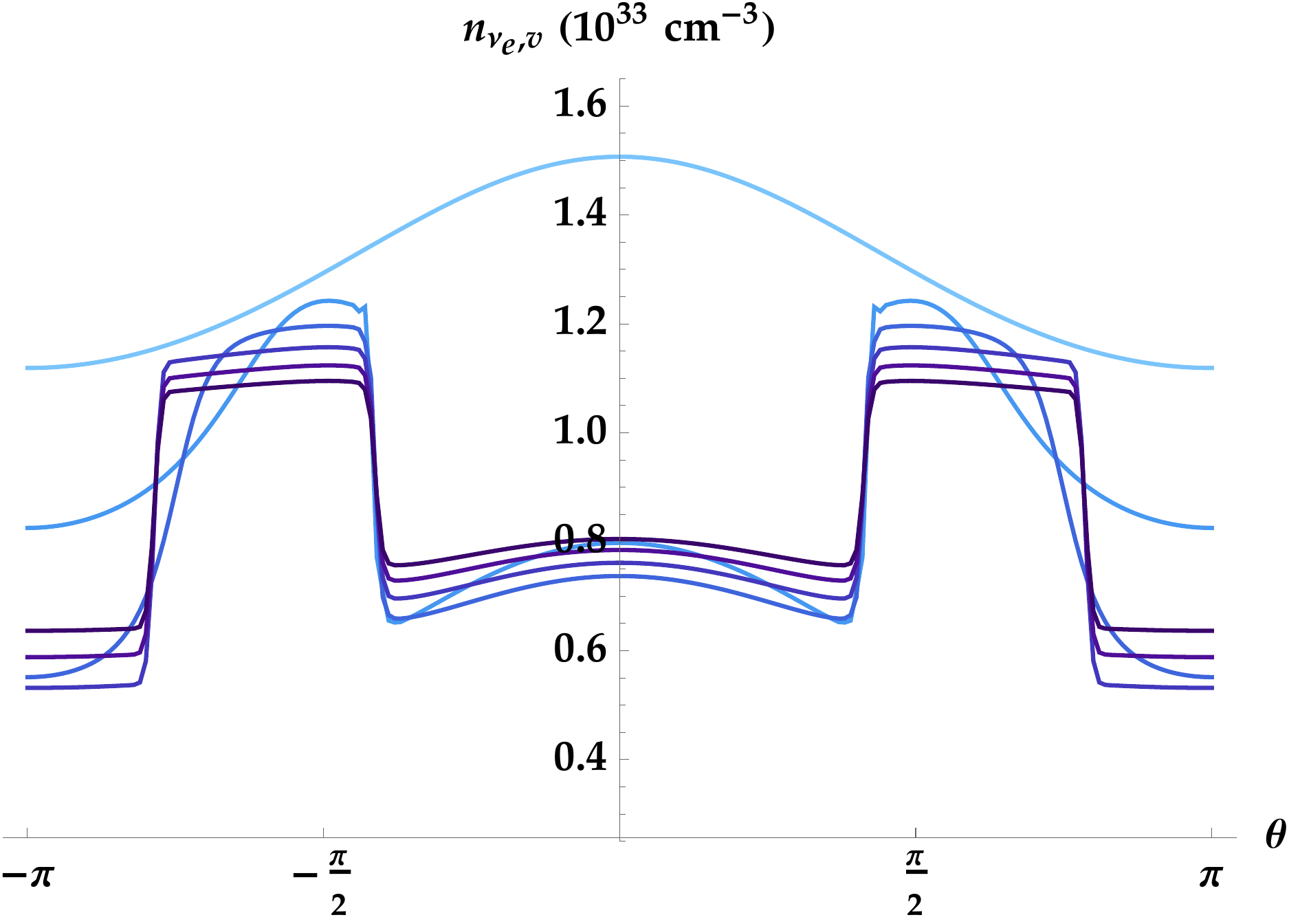}
}
\end{subfigure}
\begin{subfigure}{
\centering
\includegraphics[width=.310\textwidth]{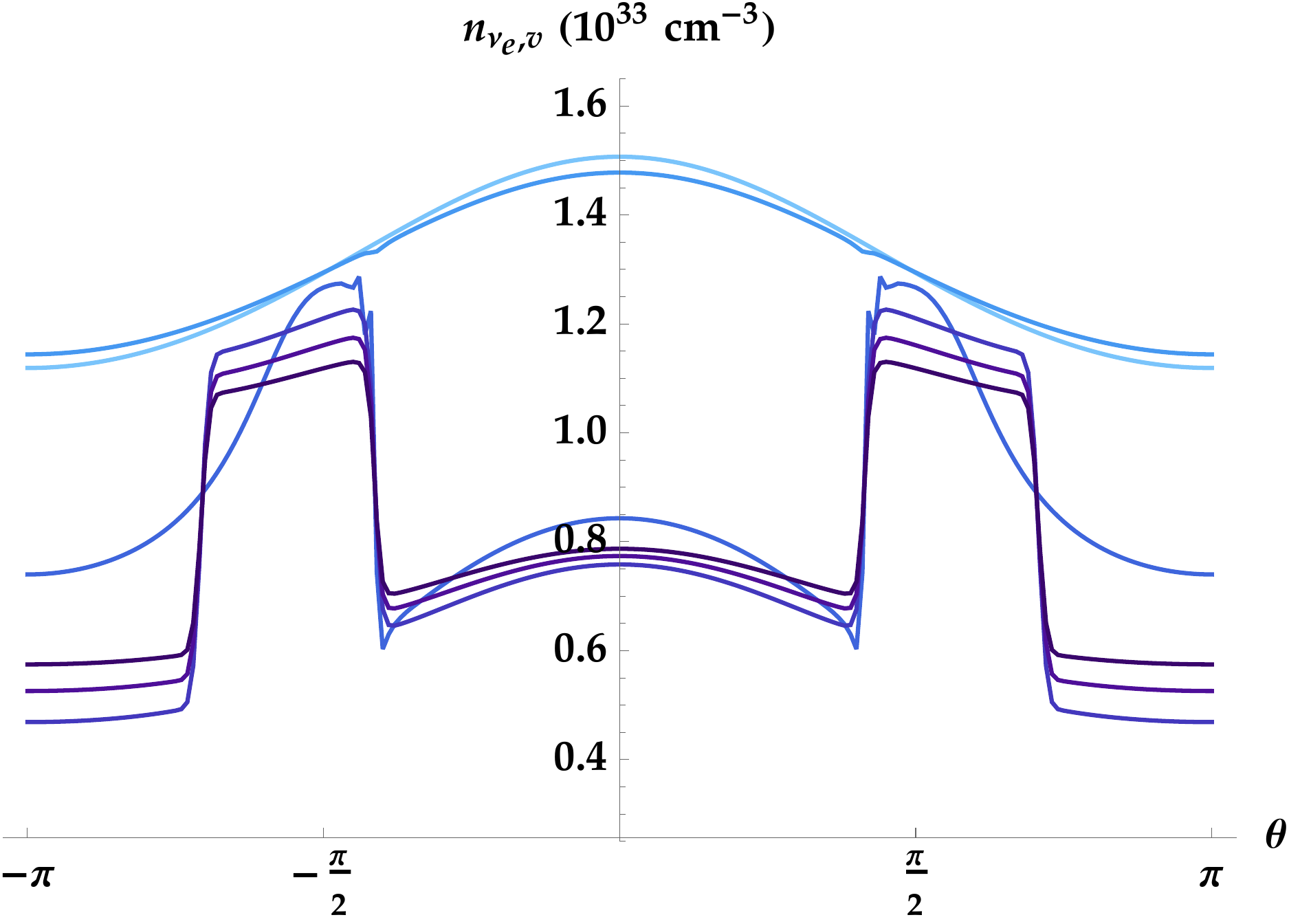}
}
\end{subfigure}
\begin{subfigure}{
\centering
\includegraphics[width=.310\textwidth]{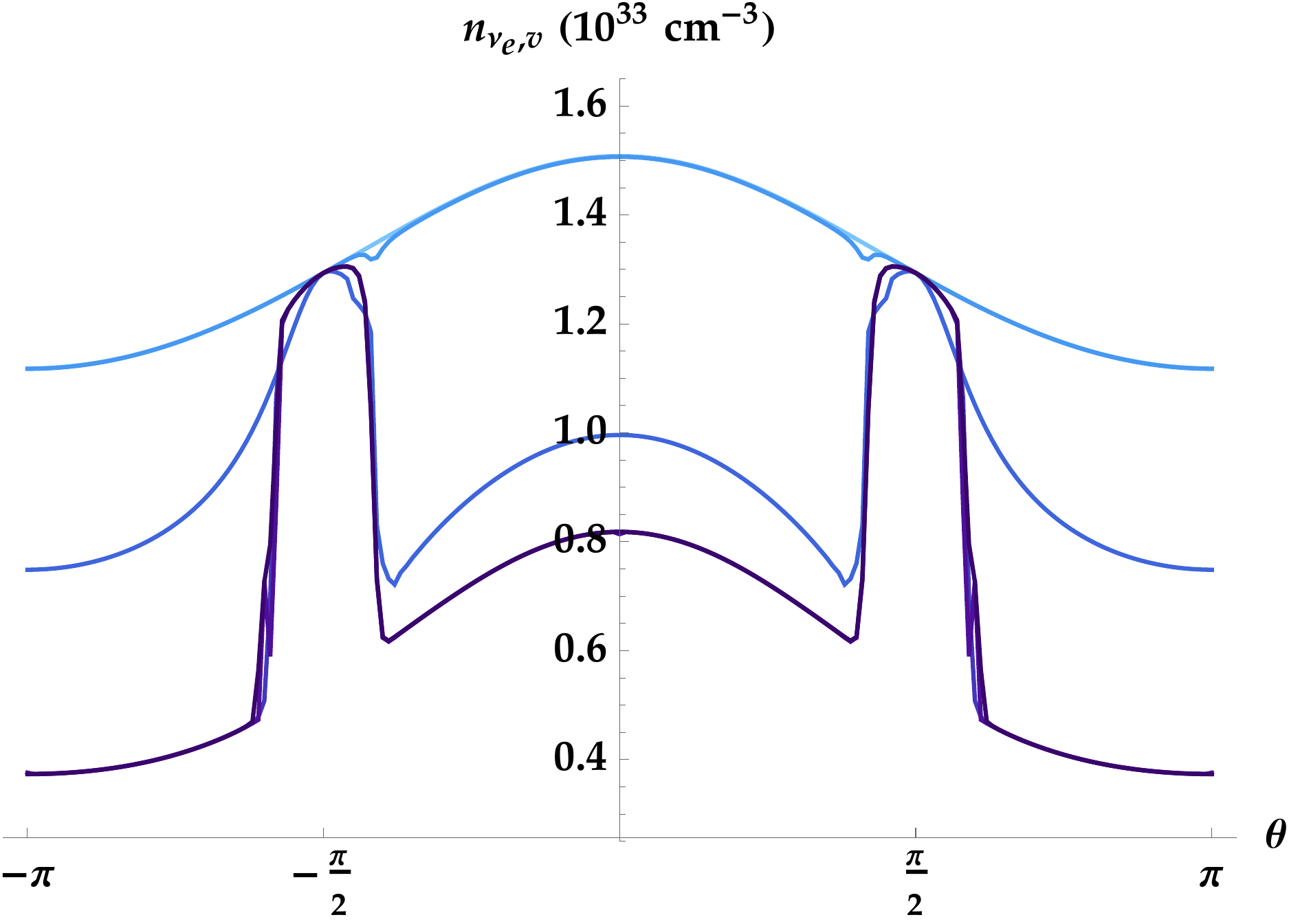}
}
\end{subfigure}
\caption{Collisionally unstable flavor evolution \textit{with} fast instability [angular distributions are given by Eq.~\eqref{eq:ang_ffc}] in the normal mass hierarchy. The evolution in the inverted hierarchy, not shown, is virtually identical. The plotted quantities are the same as in Fig.~\ref{noffc_nh}, except that in this figure the displayed snapshot times are $t = 0$, $0.06$, $0.08$, $0.10$, $0.12$, and $0.14~\mu$s for NC scattering and $t = 0$, $0.02$, $0.04$, $0.06$, $0.08$, and $0.10~\mu$s for both implementations of CC scattering. The dynamics is qualitatively similar to that seen in Fig.~\ref{noffc_nh}. The main difference is in the timing: FFC, though bringing about little flavor conversion directly, prompts the development of the collisional instability at much earlier time. FFC is visible here as the rapid, small-amplitude oscillations in the densities that start right as the densities begin to deviate from their initial values. In the collisionless version of these models, those small-amplitude oscillations would be nearly periodic, without much deviation from the initial values on the time scale shown here. See Refs.~\cite{johns2020, johns2020b, padillagay2022}, for example, for plots of comparable evolution in collisionless models. Almost all of the flavor conversion that has occurred by the final time, $t = 0.15~\mu$s, is caused by the collisional instability. The predominant effect of FFC is to shorten the growth period of the collisional instability.}  
\label{ffc}
\end{figure*}

In this section we study the evolution of systems that are unstable to both fast and collisional instabilities. We now adopt the angular distributions shown in the bottom panel of Fig.~\ref{ang_distrs}.

Results of the calculations are presented in Fig.~\ref{ffc}. Significant flavor conversion is initially instigated by the growth of the fast instability, and sets in nearly an order of magnitude faster than in the preceding section.

The magnitude of conversion from FFC itself is indicated by the depth of the first dip in $n_{\nu_e}(t)$. It is not large. The first fast oscillation period, however, gives way to the collisional instability, which produces far more flavor transformation. Other than the small-amplitude fast oscillations, which decay as the collisional instability takes over, the evolution in Fig.~\ref{ffc} is quite similar to what we observed in Figs.~\ref{noffc_nh} and \ref{noffc_ih}. The similarity speaks to the fact that we are witnessing FFC-initiated collisional instability rather than FFC processed in some other way by scattering \cite{shalgar2021, martin2021, sigl2022, kato2021, sasaki2021, hansen2022, johns2022}.

The second row of Fig.~\ref{ffc} reveals that $\mathbf{D}_1$ remains critical to the dynamics. In each case, the time elapsed during its flip, from pointing along $-\mathbf{z}$ to pointing along $+ \mathbf{z}$, is comparable to the time elapsed in the fast-stable calculations. On the other hand, because the $\mathbf{D}_1$ instability sets in much faster here, $D_1$ decays much less in the lead-up to the instability than it did in the previous figures.

The bottom two panels likewise show many of the same features as before, with differences mainly attributable to the durations of the calculations. For the NC and isotropizing CC cases, less isotropization occurs in Fig.~\ref{ffc}, which terminates at 0.15~$\mu$s, than in Figs.~\ref{noffc_nh} and \ref{noffc_ih}, which extend out to 0.4~$\mu s$. As a result, the angular features of Fig.~\ref{ffc} are less flattened out. For the non-isotropizing CC case, the shorter duration allows less time for ripples to form in the $D_{v,z} \sim 0$ region beyond $\theta = \pm \pi / 2$.

The most important point is that the waiting period prior to the visible onset of the collisional instability is greatly shortened, in agreement with Ref.~\cite{johns2021b}. Since only the onset (\textit{i.e.}, the linear growth phase) is expedited, and not the post-onset development, we infer that FFC hastens collisional instability by generating large seed coherence on a short time scale. Thus much of the linear-phase growth---which involves flavor coherence growing by several orders of magnitude---takes place at a rate proportional to $\mu | \mathbf{D}_0 |$ rather than $\Gamma$.

\section{Discussion \label{sec:discussion}}

We have presented numerical results showing collisional instabilities developing in a homogeneous, axisymmetric model. When the system is unstable to fast oscillations as well, FFC rushes the collisional instability though its linear phase. Qualitatively similar results are observed with NC and CC scattering. This effect could be important in regions that are unstable to both fast and collisional instabilities, but where the mean free path is such that collisional instabilities only grow out of the linear regime with the assistance of FFC.

Our findings are consistent with those of Ref.~\cite{johns2021b}, which claimed that FFC could expedite the development of collisional instabilities. We have expanded on the analysis there by making different collisional processes the dominant ones and by providing more numerical and analytic details on how collisional instabilities develop in anisotropic settings.

Because the model in this study combines collisions, anisotropy, and homogeneity, self-consistency is a concern \cite{johns2022}. In a realistic environment, anisotropy in the neutrino angular distributions is a product of collisions (which tend to isotropize) and inhomogeneity (which tends to make distributions more forward-peaked as a function of distance from the center of the compact-object environment). Without actually modeling the supernova or merger geometry, initially anisotropic distributions eventually isotropize. In this paper we have followed in the spirit of Ref.~\cite{johns2022} by comparing to \textit{non}-isotropizing CC scattering, to ensure that our findings are not compromised by this limitation of homogeneous models.

The anisotropic calculation of Ref.~\cite{johns2021b}, on the time scale plotted in that paper's Fig.~3, is insensitive to whether collisions are isotropizing or not. That self-consistency test was neither shown nor described, however. Here we show explicitly that the major qualitative features occur independently of isotropization and that they are therefore not artifacts of homogeneous modeling.

Recent studies have focused on the alleged enhancement of FFC by scattering. As shown in Ref.~\cite{johns2022}, this effect disappears when isotropization is taken out of the mix. Ref.~\cite{shalgar2021} presented calculations with $\Gamma = \bar{\Gamma}$ and $\Gamma \neq \bar{\Gamma}$, finding enhancement in both cases. While their $\Gamma = \bar{\Gamma}$ calculations probably show enhancement due to isotropization, the calculations with unequal NC scattering rates may show additional enhancement due to collisional instability. 

Following previous studies, we here take scattering to be the dominant process. Singling out scattering is not a realistic choice, especially when adopting highly disparate values of the neutrino and antineutrino rates, but it allows us to compare more directly to past analyses. For calculations with more realistic parameters, including contributions from different collisional processes, we refer to Ref.~\cite{johns2021b}.

In this paper we have presented new results on the development of collisional instabilities in some of the simplest models in which they occur. These results provide insights into the flavor phenomenology in scenarios where both oscillations and collisions are important. Making definitive claims about the role of collisional instabilities in real supernovae and neutron-star mergers will require more sophisticated modeling. This effort will be taken up in future work.

\begin{acknowledgments}
We thank Hiroki Nagakura for valuable conversations. L. J. was supported by NASA through the NASA Hubble Fellowship grant number HST-HF2-51461.001-A awarded by the Space Telescope Science Institute, which is operated by the Association of Universities for Research in Astronomy, Incorporated, under NASA contract NAS5-26555. Z. X. was supported by the European Research Council (ERC) under the European Union's Horizon 2020 research and innovation program (ERC Advanced Grant KILONOVA No. 885281).
\end{acknowledgments}

\bibliography{all_papers}

\end{document}